\documentclass[prd,10pt,twocolumn,nofootinbib,preprint,superscriptaddress]{revtex4}
\pdfoutput=1
\usepackage{feynmp}
\usepackage{amsmath, amssymb, amsthm, graphicx, fancyhdr,epsfig, slashed, mathrsfs}
\usepackage{pifont}% http://ctan.org/pkg/pifont
\newcommand{\cmark}{\ding{51}}%
\newcommand{\xmark}{\ding{55}}%
\usepackage{subfigure}
\usepackage{tikzsymbols}
\usepackage{natbib}
\usepackage{float}
\usepackage{lipsum}
\usepackage{mathtools} 
\usepackage{setspace}
\usepackage[normalem]{ulem}
\usepackage{tabularx}
\usepackage{tikz,xcolor,hyperref}
\hypersetup{
     colorlinks=true,
     citecolor    = blue,
     linkcolor= blue
}

\definecolor{lime}{HTML}{A6CE39}
\DeclareRobustCommand{\orcidicon}{
	\begin{tikzpicture}
	\draw[lime, fill=lime] (0,0) 
	circle [radius=0.2] 
	node[white] {{\fontfamily{qag}\selectfont \tiny ID}};
	\draw[white, fill=white] (-0.0625,0.095) 
	circle [radius=0.007];
	\end{tikzpicture}
	\hspace{-2mm}
}

\foreach \x in {A, ..., Z}{\expandafter\xdef\csname orcid\x\endcsname{\noexpand\href{https://orcid.org/\csname orcidauthor\x\endcsname}
			{\noexpand\orcidicon}}
}

\begin{document}
%========================================================
\title{Blazar Boosted ALP and vector portal Dark matter confronting light mediator searches}

\author{Sk Jeesun }
\email{skjeesun48@gmail.com}
\affiliation{School of Physical Sciences, Indian Association for the Cultivation of Science,\\ 2A $\&$ 2B Raja S.C. Mullick Road, Kolkata 700032, India}

\begin{abstract}
    The trouble in detecting low mass dark matter due to its low kinetic energy can be ameliorated in the boosted dark matter framework, where a sub-population of galactic dark matter attains very high energy after being up-scattered by energetic standard model particles.  
    However, in such a scenario the upper limits on the cross-section  obtained hitherto are typically large. 
    Hence in the minimal  extension of standard model where new mediators act as a portal between the dark and visible sectors, the direct detection limits for sub-GeV dark matter might lie within the exclusion region of other ground based searches of the mediator. 
    To evade this deadlock, we allude to blazar boosted dark matter electron scattering in multi-ton neutrino detector Super kamiokande. We consider minimal models such as axion like particle (ALP) and vector portal dark matter being upscattered by high energy blazar jet and analyse the interesting parameter reaches from Super kamiokande in the parameter space  of the mediator, surpassing the existing constraints. Besides, this scenario exhibits stronger limits for previously unexplored ALP mediated sub-MeV dark matter search which is difficult due to associated momentum suppression.  
\end{abstract}
\maketitle

\section{Introduction}
%======================================================
Numerous astrophysical and cosmological observations have evinced the presence of a non luminous non baryonic component satisfying $25\%$ of the total energy budget of the Universe coined as dark matter (DM) \cite{Zwicky:1933gu,Rubin:1970zza,PAMELA:2011bbe,Planck:2018vyg}.
The particle nature of DM persists to be an enigma for standard model (SM) of particle physics with its present set up despite the observational evidence \cite{Scherrer:1985zt,Jungman:1995df,Cirelli:2024ssz}.
To resolve this intriguing puzzle, the SM particle content is often augmented to include one or more beyond standard model (BSM) particles \cite{Arcadi:2017kky,Cirelli:2024ssz}.  
The possibility of the existence of non-zero interactions between DM and SM particles other than the gravitational one, has framed the pathway of detecting it in the direct detection experiments \cite{Feng:2010gw,Lin:2019uvt,Cirelli:2024ssz}.  
Direct detection experiments like XENONnT \cite{XENON:2022ltv,XENON:2023cxc}, LUX-ZEPLIN (LZ) \cite{LZ:2023poo,LZCollaboration:2024lux}, Panda X-II \cite{PandaX-II:2020oim}, LUX \cite{LUX:2016ggv}, DEAP-360 \cite{DEAP:2019yzn}, DarkSide-50 \cite{DarkSide:2018kuk} try to explore possible interaction of an incoming DM with nucleon or  electron in the ground based detectors while the solar system moves through the galactic DM halo \cite{Lin:2019uvt}.

All these experiments have placed strong constraints on $\mathcal{O}$(GeV) scale DM nucleon (electron) cross-sections, pushing the limits almost towards the neutrino floor \cite{LZCollaboration:2024lux}.
Despite all the heroic efforts, no direct evidence of particle DM has motivated the community to  search for low mass (sub-GeV) DM.
However, current experiments face severe challenge in detecting the low mass DM as its low velocity ($10^{-3}$) falls short of generating sufficient recoil energy above the threshold ($\sim \mathcal{O}(1)$ keV) of the ongoing detectors. 
To probe such light DM, consideration of a sub-population of it exceeding the galactic escape velocity upon scattering by high-energy cosmic ray \cite{Yin:2018yjn,Bringmann:2018cvk}, diffuse supernova neutrinos \cite{Das:2021lcr,Ghosh:2021vkt}  has  gained a lot of interests in last few years. 
In such boosted DM (BDM) scenario the galactic DM particles after being up-scattered by the homogeneous high energetic particles, are expected to scatter again with the target particles in the Earth based DM or neutrino detectors, null observations of which leads to constraint on the cross-section with nucleon (electron) \cite{Bringmann:2018cvk,Ema:2018bih,Maity:2022exk,DeRomeri:2023ytt}.
However, the probability of galactic DM being up-scattered by the cosmic particles and achieving 
high energy is very small and hence the constraints on the cross-section of sub GeV boosted DM is much smaller than those of galactic DM (with mass $\gtrsim$ GeV) \cite{Bringmann:2018cvk,Ema:2018bih}.
Recently, it has  also been observed that the BDM flux and the constraints differ significantly depending on the Lorentz structures of the interaction between DM and nucleon (or electron) \cite{Bardhan:2022bdg,Dent:2019krz,Ema:2020ulo}.
Since the cross-sections probed by such boosted DM are typically large, it is crucial to consider minimal models of the sub-GeV DM-nucleon (electron) interaction and scrutinize whether the new regions probed by direct search are allowed by the other existing constraints on the model parameters.
Novel studies in this have already investigated minimal models featuring a BSM mediator coupled to sub-GeV DM and nucleon \cite{Bell:2023sdq} or electron \cite{Guha:2024mjr}, and pointed out that the direct detection constraints (for BDM) struggle to beat the existing  constraints on the mediators coming from astrophysics and several ground based experiments.

An alternate avenue to boost light DM that has been explored recently is the upscattering of DM by blazar jet \cite{Wang:2021jic,Granelli:2022ysi}.
Blazar jet is the emitted jet of high energy particles from the active galactic nuclei (AGN) with  a supermassive black hole (BH) at the center, in close alignment to the line of sight (LOS) from Earth \cite{Urry:1995mg}.
Being a luminous source, the properties of blazars are delineated by the photon observations and are subject to modeling of the spectral energy distribution (SED) of the emitted photon \cite{Abdo_2010}.
Depending on the model, blazars can emit highly energetic flux of electrons or protons.
On the other hand, the BH at the center can accumulate a huge amount of DM particles 
leading to a spike density significantly larger than the galactic DM density \cite{Gondolo:1999ef}. Thus, the relativistic jet can up-scatter the surrounding DM particles, producing a significant amount of high energy (low mass) DM flux at Earth \cite{Wang:2021jic,Granelli:2022ysi,Bhowmick:2022zkj} \footnote{See ref.\cite{DeMarchi:2024riu,Herrera:2023nww} for related works of DM probe with blazar source.}.
Since the blazar boosted DM (BBDM) has energy larger than the threshold energy of neutrino detectors like Super Kamiokande \cite{Super-Kamiokande:2017dch}, the larger detector size and higher flux may provide stronger constraints on the DM SM cross-sections  than other boosted DM scenarios \cite{Granelli:2022ysi,Bhowmick:2022zkj}. 
Though the models of blazar spectrum  are subject to astrophysical uncertainty,  exploring minimal BBDM models with light mediators can be a legitimate exercise to gain new insights into the particle aspects of low mass DM.

Motivated by this, in this work we study low mass DM boosted by blazar jet considering a fermion DM $\chi$ coupled
to electrons ($e$) via a light mediator. As a case study, we first look into the well motivated axion like particle (ALP) $a$ as mediator.
ALP is referred to signify a broad class of pseudo-Nambu Goldstone bosons (pNGB) associated with the spontaneously broken global symmetry \cite{Peccei:1977hh}.
ALP mediated DM has been explored widely for its rich phenomenological prospects\cite{Bharucha:2022lty,Ghosh:2023tyz}.
However, detection of ALP portal sub-GeV/MeV BDM is difficult due to momentum suppression of incoming DM, leading to poor direct detection constraints \cite{Bell:2023sdq}. 
On the other hand, experiments like BaBar \cite{Bauer:2017ris}, NA64 \cite{NA64:2021aiq} place strong constraints on the ALP parameter space, making it a real challenge for the BDM detection scenario to probe any new parameter space.

We explore such ALP portal DM boosted by the energetic blazar jet in the multi-ton neutrino detector Super Kamiokande (Super-K).
Using the large fiducial volume and directional sensitivity, Super-K has the ability to set the strongest limit on such BBDM scenario \cite{Granelli:2022ysi,Bhowmick:2022zkj}.
%, as the boosted DM has energy above the threshold of Super-K ($\sim \mathcal{O}(100)$ MeV) \cite{Ema:2018bih,Granelli:2022ysi}.
As blazar source, we consider the well studied TXS 0506$+56$ \cite{Cerruti:2018tmc} and  also showcase the effect of a less distant BL Lacertae \cite{Boettcher:2013wxa} for comparison.
Existence of TXS 0506$+56$ has also been indicated by the neutrino observations at IceCube \cite{IceCube:2018dnn}, motivating the lepto-hadronic modeling of blazar jet \cite{Cerruti:2018tmc}. 
Using the benchmark model parameters and state of the art formalism, we first showcase the direct detection constraints on (sub) MeV scale ALP portal DM-$e$ interaction and a significant portion of the obtained limits go beyond the bounds set by existing ALP search experiments.
For a comparative analysis we also illustrate the constraints assuming a vector mediator.
In the later part of this paper we carry out similar exercise with  mediators that couple to both proton ($p$) and electron apart from DM, so that both of them potentially upscatter DM.
As a consequence the incoming BBDM flux increases significantly due to the higher energetic flux of blazar protons, leading to stronger limits from Super-K. 

We orgnaize the paper as follows. In Sec.\ref{sec:bbdm_flux} we present the formalism of obtaining BBDM flux. Then we discuss the results for electrophilic DM in Sec.\ref{sec:electro} and DM with both nucleon and electron coupling in Sec.\ref{sec:nucleo}.
Finally we conclude in Sec.\ref{sec:conc}.
Some relevent calculations related to BBDM flux is shown in Appendix \ref{sec:energy}-\ref{apx:dm_density}.
A detail discussion on the existing constraints on the BSM mediator has been put in Appendix \ref{apx:constraint}.

\section{Blazar boosted DM flux}
\label{sec:bbdm_flux}
%===================================================
\subsection{Flux of Blazar jet electrons:}
\label{sec:spec}
%======================
The blazar jet is evaluated considering the popular ``blob geometry", assuming the particles are emitted isotropically from a homogeneous blob in the jet \cite{Wang:2021jic}.
The blob moves with velocity $\beta_B$ along the jet axis with respect to the rest frame of the BH center. 
For an observer in rest with respect to the BH, the corresponding Lorentz boost factor is given by, $\Gamma_B=\sqrt{1-\beta_B^2}$.
The angle of inclination of jet axis with respect to the line of sight (LOS) of the observer is defined as $\theta_{LOS}$.

For a given (lepto-)hadronic model, the energy spectrum of the high-energetic particles in the blob frame is provided by a power law distribution \cite{Cerruti:2020lfj},
\begin{eqnarray}
    \frac{d\Gamma_i'}{dE_i' d\Omega'}=\frac{1}{4\pi} c_i \left(E_i'/m_i \right)^{-\alpha_i},
\end{eqnarray}
where $E_i,m_i$ are the energy and mass of  the emitted particle $i\in\{e,p\}$ respectively, and the ``prime" in the superscript signifies the quantities to be in blob frame.
$c_i$ is the normalization constant computed from the predicted electron/proton luminosity $L_i$ (shown in eq.\eqref{eq:ce}).  Quantities without a ``prime" in the superscript refer them to be in the observer's rest frame. 
Here we focus on the electron jet ($i\equiv e$) which is predicted to be an essential component of blazar spectrum by several models of photon SED of blazars \cite{Ghisellini_2009,Cerruti:2020lfj,Celotti:2007rb}.

To obtain the boosted DM flux, one needs the blazar jet spectrum of electrons in the observer's frame given as (using eq.\eqref{eq:trans})
\begin{eqnarray}\nonumber
    \frac{d\Gamma_e}{dE_e d\Omega}=&&\frac{1}{4\pi} c_e \left(E_e/m_e \right)^{-\alpha_e}\\
    &&\times \dfrac{\beta_j (1-\beta_j \beta_B \mu)^{-\alpha_e}\Gamma_B^{-\alpha_e}}{\sqrt{(1-\beta_j\beta_B\mu)^2-(1-\beta_j^2)(1-\beta_B^2)}},
    \label{eq:spectrum}
\end{eqnarray}
where, $E'$ has been replaced by $E$ using eq.\eqref{eq:gam}. We denote the kinetic energy (KE) of the electron as $T_e=E_e-m_e$.
$\mu=\cos{\theta}$, where $\theta$ is the angle with respect to jet axis in the observer's frame.
$\beta_{j},j=e$ is the speed of emitted particle given by, $\beta_j=\sqrt{1-\gamma_e^2}$, $\gamma_e$ being $E_e/m_e$ in rest frame.
In blob frame $E'/m_e$ has two cut-offs in both high and low values i.e. $\gamma'_{\rm max}$ and $\gamma'_{\rm min}$.
For a particular SED $\gamma'_{\rm max}$, $\gamma'_{\rm min}$, $\alpha_e$, $L_e$ and the Doppler factor $\mathcal{D}= \Gamma_B^{-1} (1-\cos{\theta_{\rm LOS}})^{-1}$ are fitted to obtain these model parameters.
For this work we consider the model parameters for TXS 0506$+56$ and BL Lacertae as pointed out in the existing literature \cite{Granelli:2022ysi}.
Aforesaid five parameters along with $\theta_{LOS}$, the mass of black hole ($M_{\rm BH}$) in the AGN center , luminosity distance ($d_L$) and redshift ($z$) are tabulated in table \ref{tab:1}.
Throughout this paper, in all the tables and figures we denote TXS 0506$+56$ (BL Lacertae) as ``TXS" (``BLL").

\begin{table}[]
%\begin{center}
\centering
\begin{tabular}{ c c c }
Parameter~~~~ & TXS$0506+56$~~~ & BL Lacertae \\
 \hline \hline
 $L_e ({\rm erg}/s)$ & $1.32 \times 10^{44}$ & $8.7 \times 10^{42}$ \\  
 $(\gamma'_{\rm min}, \gamma'_{\rm max})$ & ($500, 1.3 \times 10^4$) & ($700, 1.5 \times 10^4$) \\
 $\mathcal{D}$  & 40 & 15 \\
  $\alpha_e$  & 2.0 & 3.5 \\
  $\theta_{\rm LOS}(^\circ)$  & 0 & 3.82 \\
  $d_L({\rm Mpc})$  & 1835.4 & 322.7 \\
  $z$  & 0.337 & 0.069 \\
  $M_{\rm BH} (M_\odot)$  & $3.09\times 10^8$ & $8.65\times 10^7$ \\
 \hline
\end{tabular}
\caption{Model parameters for the blazar source TXS 0506$+56$ \cite{Cerruti:2018tmc} and BL Lacertae \cite{Boettcher:2013wxa}.}
 \label{tab:1}
%\end{center}
\end{table}

%===============================
\subsection{DM density profile}
%===================
If the super massive black hole (SMBH) at the centre of galaxy grows adiabatically, it is expected that a significant amount of DM particles would accumulate around it \cite{Gondolo:1999ef}.
This type of enhanced DM density is referred as DM spike in the literature
\cite{Gorchtein:2010xa}.
The DM density $\rho(r)$ follows a  power law grading $\rho_1(r)\propto r^{-\gamma}$ within the spike radius $R_{sp}$, where, $\gamma$ is the power of the density grading and governs the depletion of spike density with distance $r$ from the center \cite{Gondolo:1999ef}. 
Outside the spike DM density follows as usual the NFW profile
$\rho_2(r)\propto r^{-1}$.
Generally the spike radius is taken to be $R_{sp}=10^5 R_s$ \cite{Wang:2021jic,Granelli:2022ysi}, where $R_s=2 GM_{\rm BH}/c^2~(c=1$ in natural unit) is the Schwarzschild radius \cite{Gondolo:1999ef,Gorchtein:2010xa}. 
The DM density within the distance $4 R_s$ from the galaxy center is expected to accrete and hence the density $\rho_1 (r<4R_s)=0$.
It is customary in literature to normalize the total DM mass within $R_{sp}$ to be equal to $M_{\rm BH}$, i.e $\int_{4R_s}^{R_{sp}} d^3r \rho(r)=M_{\rm BH}$.
The normalization outside $R_{sp}$ is obtained using the continuity at $R_{sp}$ i.e. $\rho_2(R_{sp})=\rho_1(R_{sp})$.

So far what we discussed is valid for non-annihilating DM.
When DM particles annihilate, they may deplete the DM number density throughout the time scale. 
%DM with number density $n_\chi$ and thermal averaged annihilation cross section $\langle \sigma v \rangle$ will reduce its density at a rate $\Gamma = n_\chi \langle \sigma v \rangle$. 
%After time $t$ the maximum surviving DM density is given by $\rho(t)=\frac{\rho(t=0)}{\Gamma t}=m_\chi/\langle \sigma v \rangle t$.
This will flatten the spike density and after time $t_{\rm BH}$ i.e. the age of black hole, the density will be \cite{Gondolo:1999ef},
%a core density which will be shown shortly. 
%Using $\partial \rho /\partial t=- \langle \sigma v \rangle  \rho^2/m_\chi$, and integrating from $t=0$ to $t_{\rm BH}$, the age of black hole one obtains \cite{Gondolo:1999ef},
\begin{eqnarray}
    \rho(r,t_{\rm BH})= \dfrac{\rho(r,t=0) \rho_{\rm core}}{\rho(r,t=0) +\rho_{\rm core}} ,
\end{eqnarray}
where $ \rho_{\rm core}=m_\chi/\langle \sigma v \rangle t_{\rm BH}$ is the maximum surviving density after $t_{\rm BH}$, often known as saturation density in literature \cite{Gondolo:1999ef}. 
$\langle \sigma v \rangle$ is the thermal averaged annihilation cross section. 
For non annihilating DM, $\langle \sigma v \rangle=0$, $\rho(r,t_{\rm BH})= \rho(r,t=0)$.
Variation of the DM density profile with the distance from BH center (in terms of $R_s$) has been portrayed in Appendix \ref{apx:dm_density}.

Another important quantity to determine the expected boosted flux is the line of sight integral of DM density defined as $\Sigma_{\rm DM} (r)=\int_{4R_s}^{r} \rho(r')~ dr'$.
$\Sigma_{\rm DM} (r)$ saturates at $r\gtrsim 10^5 R_s$ and hence $\Sigma_{\rm DM}^{\rm tot} =\Sigma_{\rm DM} (10^5 R_s)$ is used to define the total LOS integral as conventionally  done in literature \cite{Wang:2021jic}.

There are different values of $\gamma$ found in the literature, which may result in different density profiles \cite{Gondolo:1999ef,Ullio:2001fb}.
The variation of $\Sigma_{\rm DM}^{\rm tot}$ with $\gamma$ is discussed in Appendix \ref{apx:dm_density}.
For a comparative analysis with existing literature \cite{Wang:2021jic,Bardhan:2022bdg}, in this work we pick two well studied profiles of the spike:  $\gamma=7/3$ \cite{Gondolo:1999ef} and $\gamma=3/2$ \cite{Gnedin:2003rj}.
The qualitative change in the direct search limits for other values of $\gamma$ can be easily comprehended  
from the discussion in Appendix \ref{apx:dm_density}.
Having a detailed discussion about the blazar jet and DM spike density we are now set to evaluate the boosted DM flux.

%=================================
\subsection{Flux of Boosted DM}
As mentioned earlier, the high energy electrons from the blazar jet may up-scatter the surrounding DM particles. 
Since the DM particles have very small velocity $v\sim10^{-3}c$ compared to blazar jet, DM can be treated as at rest.
Upon scattering the differential flux of the incoming DM is given by \cite{Wang:2021jic},
\begin{eqnarray}
    \frac{d\phi_\chi}{dT_\chi}= \frac{\Sigma_{\rm DM}^{\rm tot}}{2 \pi m_\chi d_L^2} \int_0^{2\pi} d\phi_s \int_{T_e^{\rm min}}^{T_e^{\rm max}} dT_e \frac{d\sigma_{\chi e}}{dT_\chi} \frac{d\Gamma_e}{dT_e d\Omega_e}.
    \label{eq:flux}
\end{eqnarray}
$\phi_s$ is the azimuthal angle with respect to LOS.
%Since the blazar axis is aligned by an angle $\theta_{\rm LOS}$ with respect to line of sight, 
The cosine of the polar angle in the observer frame ($\mu$) is related to
the cosine of the scattering angle ($\bar{\mu}$) corresponding to DM rest frame
by a 2 dimensional rotation of angle $\theta_{LOS}$ given as \cite{Wang:2021jic},
\begin{eqnarray}
    \mu(\bar{\mu},\phi_s)= \bar{\mu} \cos{\theta_{\rm LOS}}+ \sin{\phi_s}  \sin{\theta_{\rm LOS}}\sqrt{1-\bar{\mu}^2},
\end{eqnarray}

where, $\bar{\mu}$ is defined as \cite{Wang:2021jic},
\begin{eqnarray}
    \bar{\mu} (T_e,T_\chi)&=&\Bigg[1+\dfrac{T_\chi^{\rm max}-T_\chi}{T_\chi}  \dfrac{(m_\chi+m_e)^2+2 m_\chi T_e}{(m_\chi+m_e+T_e)^2}\Bigg]^{-1/2}.
\end{eqnarray}
$T_\chi^{\rm max}=(T_e^2+2 m_e T_e)(T_e+(m_\chi+m_e^2)/(2 m_\chi))^{-1}$ is the maximum kinetic energy obtained by DM upon scattering with $e$ and $\mu_{\chi e}$ is the reduced mass of DM electron system.

Coming back to eq.\eqref{eq:flux}, 
$T_e^{\rm min}(T_\chi)$ is the minimum kinetic energy required to up-scatter a non-relativistic DM to energy $T_\chi$ and is given by \cite{Bringmann:2018cvk,Bardhan:2022bdg},
\begin{equation}
    T_{e}^{\rm min} =
    \left(\frac{T_\chi}{2}-m_e \right) \left[1 \pm \sqrt{1+\frac{2 T_\chi (m_e+m_\chi)^2}{m_\chi(2 m_e-T_\chi)^2}}\right] ,
    \label{eq:tmin}
\end{equation}
where $(+)$ or $(-)$ sign is applicable when $T_\chi > 2m_e$ or $T_\chi < 2m_e$, respectively. However, according to the SED modeling \cite{Cerruti:2018tmc, Boettcher:2013wxa}, the blazar jet itself has a cut off in the low energy $E'_{e,\rm min}=m_e\gamma'_{e,\rm min}$. Hence the lower cut off in the kinetic energy of of jet electrons in rest frame is given by, $T_{e,\rm jet}^{\rm min}=m_e(\gamma'_{e,\rm min}\Gamma_B^{-1}(1-\beta_B \beta \cos\theta_{\rm LOS})^{-1}-1)$.
Thus the lower limit of the above mentioned integration becomes Max$\{T_{e}^{\rm min}, T_{e,\rm jet}^{\rm min}\}$.
On the other hand, the upper cut off in the energy of the blazar jet sets the upper  limit $T_{e,\rm jet}^{\rm max}=m_e(\gamma'_{e,\rm max}\Gamma_B^{-1}(1-\beta_B \beta \cos\theta_{\rm LOS})^{-1}-1)$.
Finally, $d\sigma_{\chi e}/dT_\chi$ is the differential cross section of DM and incoming $e$.
While computing $d\sigma_{\chi e}/dT_\chi$ the information about the underlying particle physics model is important and will be discussed in the following sections.

Being upscattered DM particles may reach Earth and the incoming high energy DM particles are expected to scatter with target electrons in the multi-ton neutrino detectors.
The expected differential recoil rate is given by,
\begin{equation}
    \frac{dR}{dE_R}=t_{\rm exp} N_{\rm T} \mathcal{E}(E_R) \int^{\infty}_{T_{\chi}^{\rm min}(E_R)} \frac{d\Phi_\chi^{\rm tot}}{dT_\chi} \frac{d \sigma_{\chi e}}{dE_R}dT_{\chi},
    \label{eq:recoil}
\end{equation}
%==========
where $E_R$ is the recoil energy, $t_{\rm exp}$
is the time of exposure and $N_T$ denotes the number of 
target particles. $T_{\chi}^{\rm min}$ is the minimum kinetic energy 
of an incoming DM particle to produce a recoil energy $E_R$ which is given by
\cite{Bringmann:2018cvk,Bardhan:2022bdg},
\begin{equation}
    T_{\chi}^{\rm min}=
    \left(\frac{E_R}{2}-m_\chi \right) \left[1 \pm \sqrt{1+\frac{2 E_R (m_e+m_\chi)^2}{m_e(2 m_\chi-E_R)^2}} \right] ,
    \label{eq:tmin2}
\end{equation}
where $(+)$ or $(-)$ sign is applicable when $E_R > 2m_\chi$ or $E_R < 2 m_\chi$, respectively.
$\mathcal{E}(E_R)$ is the efficiency factor of the experiment.
$d \sigma_{\chi e}/dE_R$ is the differential cross-section of incoming energetic $\chi$ and target $e$ at rest.

\section{Electrophilic DM}
\label{sec:electro}
%===========================
To evaluate the BBDM flux and expected DM signature, one needs the information about the Lorentz structure of the presumed interaction between DM and SM.
Thus, it is crucial to consider at least a minimal effective DM model. 
In this section, we consider DM models that can allow interaction of DM with only $e$, which we refer as electrophilic DM.
%===========================
\subsection{ALP portal}
We consider a minimum DM scenario coupled to a pseudo scalar dubbed as ALP $a$ like,
\begin{equation}
    \mathcal{L}\supset i g_{a\chi} a \bar{\chi} \gamma^5 \chi + i g_{ae} a \bar{e} \gamma^5 e,
\end{equation}
where, $g_{a \chi}$ and $g_{ae}$ are the ALP-DM and ALP-electron couplings respectively.
Thus $\chi$ can scatter with ambient jet $e$ through the ALP portal.
The differential elastic cross section between $\chi$ and incoming $e$ is given by,
\begin{eqnarray}
    \frac{d\sigma_{\chi e}}{dT_\chi}= \frac{1}{8\pi} \frac{g_{a \chi}^2 g_{a e}^2 m_\chi T_\chi^2}{(2 m_e T_e+T_e^2)(m_a^2+2 m_\chi T_\chi)^2},
    \label{eq:boost_a}
\end{eqnarray}

\begin{figure}
    \centering
    \subfigure[\label{d1}]{
    \includegraphics[scale=0.4]{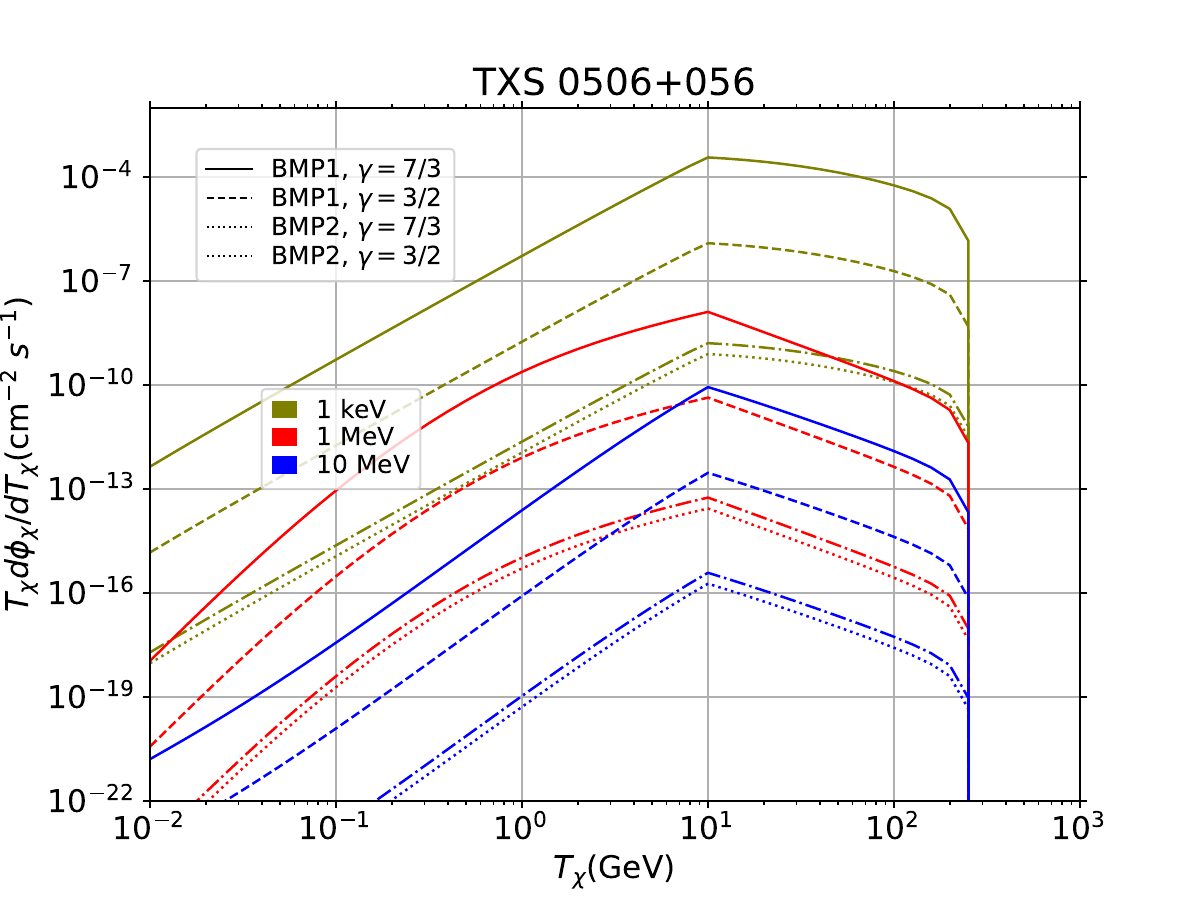}}
    \subfigure[\label{d2}]{
    \includegraphics[scale=0.4]{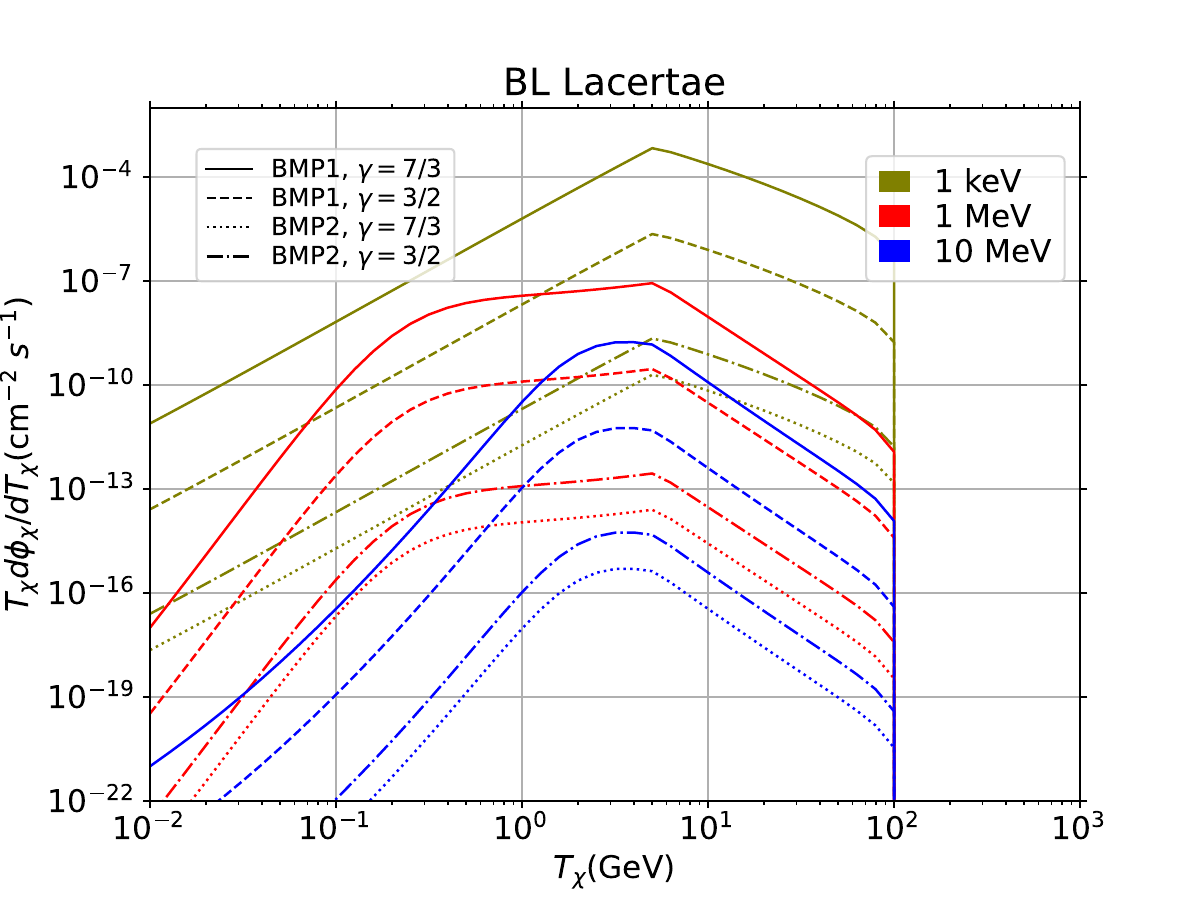}}
    \caption{Variation of differential flux of BBDM with $T_\chi$  for fixed values of $g_\chi g_e=10^{-4},~m_a=100$ MeV for (a) TXS 0506$+56$ and (b) BL Lacertae. The flux due to three different DM masses $m_\chi=1$ keV, 1 MeV ,10 MeV are shown by olive, red and blue color respectively. The solid and dashed lines signify BMP1 with $\gamma=7/3$ and $3/2$ respectively,
whereas, the dashed dot and dotted lines signify BMP2 with $\gamma=7/3$ and $3/2$ respectively.}
    \label{fig:dmflux}
\end{figure}

In Fig.\ref{fig:dmflux} we present the variation of differential flux of boosted DM using eq.\eqref{eq:flux} with $T_\chi$ for TXS 0506$+56$ (Fig.\ref{d1}) and BL Lacertae (Fig.\ref{d2}) for fixed values of $g_\chi g_e=10^{-4},~m_a=100$ MeV.
we show the flux for three different DM masses $m_\chi=1$ keV, 1 MeV ,10 MeV shown by olive, red and blue colors respectively.  
Also we consider two benchmark points (BMP) for this plot: $\langle \sigma v\rangle = 0$ (BMP1) and $\langle \sigma v\rangle = 3\times 10^{-26}$ cm$^3$s$^{-1}$ (BMP2).
The solid and dashed lines signify BMP1 with $\gamma=7/3$ and $3/2$ respectively.
On the other hand dashed dot and dotted lines signify BMP2 with $\gamma=7/3$ and $3/2$ respectively.
As expected with lower $m_\chi$ DM number density increases, leading to higher DM flux and
for both the blazar sources BMP1 results in higher DM flux for a fixed $\gamma$.
The nonzero annihilation cross-section for BMP2 leads to a decrease in the incoming DM flux.
For same BMP $\gamma=7/3$ results in higher flux than $\gamma=3/2$ for higher LOS density (see Appendix \ref{apx:dm_density}).
The sharp cut-offs in high $T_\chi$ for both the blazar sources are the consequence of the upper limit on the blazar jet electron energy $T_{e,\rm jet}^{\rm max}$.
The higher value of $\gamma'_{e,\rm max}$ for TXS 0506$+56$ (Fig.\ref{d1}) results in higher flux for $T_\chi \gtrsim 100$ GeV which will lead to more stringent constraints on DM parameter space as will be shown in the following discussion.

%\section{Results}
%=================================================

To evaluate the recoil rate in the detector we use eq.\eqref{eq:recoil} with $d \sigma_{\chi e}/dE_R$  given by,
\begin{eqnarray}
    \frac{d\sigma_{\chi e}}{dE_R}= \frac{1}{8\pi} \frac{g_{a \chi}^2 g_{a e}^2 m_e E_R^2}{(2 m_\chi T_\chi+T_\chi^2)(m_a^2+2 m_e E_R)^2}.
\end{eqnarray}

For this work we consider Super Kamiokande (Super-K) as detector due to its larger target size $N_T=7.5 \times 10^{33}$ with exposure time $2628.1$ days \cite{Super-Kamiokande:2017dch}. 
We evaluate  the DM signal  in $3$ different energy bins: 0.1-1.33 GeV (Bin 1), 1.33-20 GeV (Bin 2), 20-$10^3$ GeV (Bin 3) following the approach developed in ref.\cite{Granelli:2022ysi}.
For each bin the event rate , expected background and signal efficiency are provided in ref.\cite{Super-Kamiokande:2017dch}.
Using the angular information of the incoming blazar jet the dominant atmospheric neutrino background can be vetoed. 
We use the statistical analysis performed in ref.\cite{Granelli:2022ysi} 
 to draw the $95\%$ confidence level (CL) upper limit for ALP mediated boosted DM electron cross-section.

\begin{figure}[!tbh]
    \centering
    \subfigure[\label{c1}]{
    \includegraphics[scale=0.45]{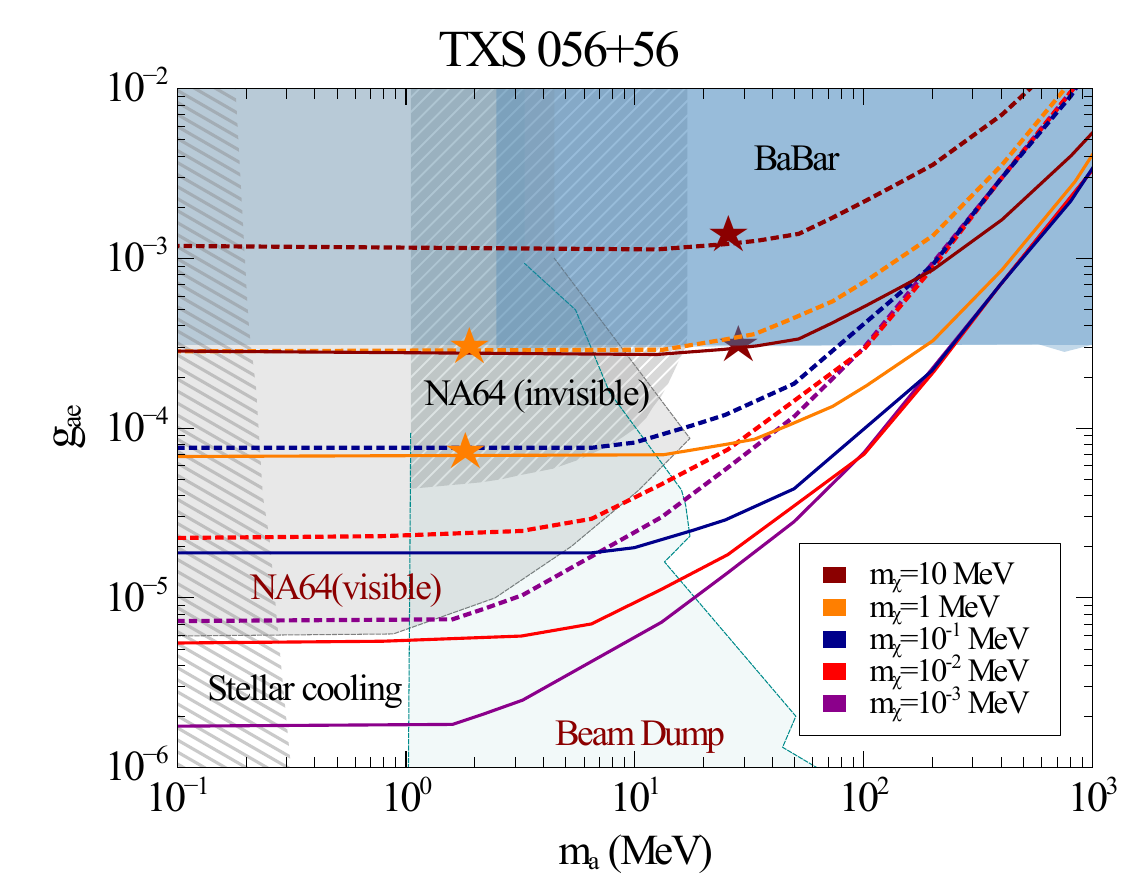}}
    \subfigure[\label{c3}]{
    \includegraphics[scale=0.45]{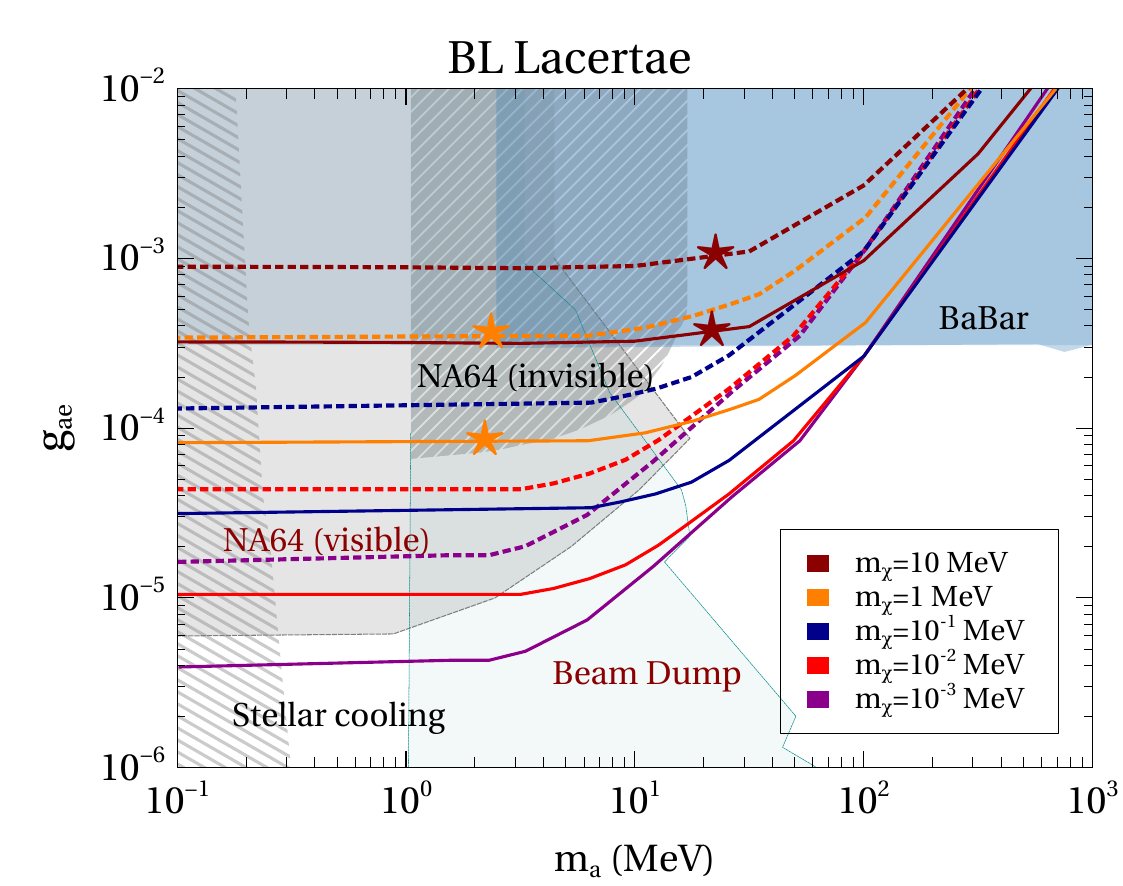}}
    \caption{$95\%$ CL Constraints on BBDM in $m_a$ vs. $g_{ae}$ parameter space along with other existing constraints on ALP considering (a) TXS 056+56 and (b) BL Lacertae as blazar source.  Regions above the individual colored lines are excluded from DM searches for different DM mass $m_\chi=$10 MeV, 1 MeV, 0.1 MeV, $10^{-2}$ MeV, $10^{-3}$ MeV denoted by  brown, orange, blue, red and magenta colored solid (dashed) lines for $\gamma=7/3$ ($\gamma=3/2$) respectively. The different shaded regions correspond to existing constraints on ALP discussed in the text. The colored ``star" marks signify the lower limit on $m_a$ of the BaBar constraint ($g_{ae}<3\times 10^{-4}$) for different DM masses. }
    \label{fig:constraints}
\end{figure}

Before delving into the results from Super-K we highlight the existing constraints on the light electrophilic mediator. 
A brief summary of existing experimental constraints on ALP mediator has been made in Appendix \ref{apx:constraint}.
%Among them, the most relevant constraints come from BaBar \cite{BaBar:2014zli,Bauer:2017ris,Liu:2023bby,Armando:2023zwz,Angel:2023exb} and NA64 experiment (visible and invisible decay of the mediator) \cite{NA64:2021aiq,Andreev:2021fzd}.
%
To show the direct search limits on $g_{ae}$ (Fig.\ref{fig:constraints}) in our analysis we consider  the optimistic benchmark with $g_{a\chi}=\sqrt{4\pi}$ \cite{Bell:2023sdq}.
Thus when $m_a>2 m_\chi$, ALP decays to dark sector with  almost $\sim 100\%$ BR leading to very tiny room for visible decays and hence the visible decay constraints become negligible in that regime. 
This point is crucial in understanding our limits on BBDM from Super-K in Fig.\ref{fig:constraints}. 

%\begin{widetext}
%\end{widetext}

In Fig.\ref{fig:constraints} we present the obtained limits on the blazar boosted DM in the mass vs. coupling parameter space of ALP mediator.
%\cite{Knapen:2017xzo}.
For comparison we display the constraints from TXS 0506$+56$ (Fig.\ref{c1}) and BL Lacertae (Fig.\ref{c3}).
Constraints with $\gamma=7/3$ ($\gamma=3/2$) are depicted by solid (dashed) lines.
The regions above individual colored lines are excluded from DM searches for different DM mass $m_\chi=$10 MeV, 1 MeV, 0.1 MeV, $10^{-2}$ MeV, $10^{-3}$ MeV denoted by  brown, orange, blue, red and magenta colors respectively.
We show the constraints assuming $\langle\sigma v\rangle =0$ for both the blazar sources to obtain an optimistic limit.
In the same parameter space we also showcase current constraints from different searches of electrophilic ALP.
For larger couplings constraints from BaBar  \cite{Bauer:2017ris,Liu:2023bby,Armando:2023zwz,Angel:2023exb} is relevant shown by light blue shaded region. For ALP mass $\sim \mathcal{O}(1)$MeV constraints from different Beam Dump experiments (light cyan shaded region) \cite{Bechis:1979kp,Blumlein:1991xh,Andreas:2010ms,Liu:2017htz,Waites:2022tov}, 
%\bl{NA48/2 (from K-meson decay to $e^+e^-$) \cite{NA482:2009pfe,Bharucha:2022lty}}
%for $1 $ MeV$\lesssim m_a\lesssim 15$ MeV 
and NA64 invisible decay  (solid grey region) \cite{NA64:2021aiq,Andreev:2021fzd} set the most stringent limit.
In the MeV-sub MeV region NA64 visible decay constraint  \cite{NA64:2021aiq,Andreev:2021fzd} is shown by the backward diagonal grey region.
Astrophysical constraints like stellar cooling is relevant in the sub-MeV region shown by the forward diagonal light gray region \cite{Caputo:2024oqc}.

%\textcolor{red}{Here we highlight some key aspects of the existing constraints. Readers not intrigued by the laboratory bounds on ALP may skip this paragraph. 
For comparative analysis we plot all the limits for different $m_\chi$ in the same plane, although some of the existing constraints change with variation in DM mass as hinted earlier. 
The summary of relevant constraints for 
different DM masses are presented  in Table \ref{tab:2}.
For $m_\chi=10$ MeV (brown line) and $m_\chi=1$ MeV (orange line) BaBar excludes $g_{ae}>3\times 10^{-4}$ from $m_a\sim15$ GeV to only $m_a=20$ MeV and $m_a=2$ MeV respectively, shown by the colored ``star mark" on the individual limits. 
So, practically there is no limit from BaBar on the left side of the ``star mark" for respective $m_\chi$.
%For $m_\chi=10^{-1}$ MeV (blue line) the exclusion region from BaBar extends down to  $m_a=0.2$ MeV, however that region is already excluded by stellar cooling.
Similarly, the NA64 invisible decay constraint doesn't apply for $m_\chi=10$ MeV.
%On the other hand NA64 visible decay constraint starting from $m_a\sim 10$ MeV does not apply for $m_\chi \in 10^{-1},10^{-2},10^{-3}$ MeV.
On the other hand, ``Beam Dump" constraints (visible) are applicable for only $m_\chi=10$ MeV for our region of interest and not for other chosen DM masses.
%i.e. $m_\chi \in 1,10^{-1},10^{-2},10^{-3}$ MeV.
To signify the relevance of NA64 visible decay and Beam Dump constraints for only $m_\chi=10$ MeV we show the labels in brown. 
%\textcolor{magenta}{Despite the  existing experimental constraints, our BBDM limits do explore new parameter reaches for MeV (-sub MeV) DM compared to cosmic ray boosted DM.}
%==================================
\begin{widetext} 
\begin{table*}[!tbh]
%\begin{center}
\centering
\begin{tabular}{ c c c c c c c}
$m_\chi$(MeV)~ & ~~BaBar(MeV)~~~ &~~NA64 visible ~~& ~NA64 invisible~~& Beam Dump~~& ~~TXS ~~~&BLL\\
 \hline \hline
 $10$ & $20$-$15\times 10^3$  & $m_a<20$ MeV& \xmark & \cmark & 12-20 MeV& 12-20 MeV\\  
 $1$ & $2$ -$15\times 10^3$ & $m_a<2$ MeV & $m_a>2$ MeV &\xmark &  30-80MeV &  40-60MeV\\ 
 $10^{-1}$  & $0.2-15\times 10^3$  & \xmark & \cmark &\xmark &0.3-110 MeV& 0.3-90 MeV \\
  $10^{-2}$  & \cmark & \xmark & \cmark &\xmark & 0.3-110 MeV& 0.3-90 MeV\\
  $10^{-3}$  & \cmark & \xmark & \cmark &\xmark & 0.3-110 MeV & 0.3-90 MeV \\
 \hline
\end{tabular}
\caption{Possible constraints and their range on $m_a$ for different $m_\chi$. The last two coloumns named ``TXS" and ``BLL" represent the range of $m_a$ of the ``new" parameter space probed by BBDM searches at Super-K assuming TXS 0506$+56$ (Fig.\ref{c1}) and BL Lacertae (Fig.\ref{c3})  as blazar source  with $\gamma=7/3$. Ticks refer to the constraints that are applicable to all the regions as shown in the plot. Cross-mark signifies that the respective constraint is not applicable.}
 \label{tab:2}
%\end{center}
\end{table*}
\end{widetext}

%===================================
The strongest bound for ALP mediated DM electron cross-section comes from TXS 0506$+56$ (Fig.\ref{c1}) with $\gamma=7/3$.
%BMP1 leads to more stringent bound than BMP2 as the DM spike density decreases for the letter one.
As discussed in context of Fig.\ref{fig:dmflux}, $\gamma=3/2$ leads to smaller DM flux than with $\gamma=7/3$ for both the blazar sources, 
the bounds are more stringent for $\gamma=7/3$.
%On the other hand for BMP2 the flux is slightly larger with for $\gamma=3/2$ than with $\gamma=7/3$ leading to slightly better sensitivity.
For, $m_a\sim 30-100$ MeV, $g_{ae}\lesssim 10^{-4}$ the parameter space remains allowed by all other existing searches. 
Considering ALP mediated DM, Super Kamiokande excludes these parameter spaces for $m_{\chi}\lesssim 10$ MeV.
With decreasing DM mass the enhanced DM flux leads to more stringent constraint. 
For larger DM masses ($m_\chi\gtrsim 10$ MeV) the bounds obtained are already within existing constraints for $m_a\gtrsim m_\chi$ and hence can't probe any new parameter space. 
For even lighter DMs ($m_\chi \leq 1$ MeV) the limits on the ALP mass-coupling parameter space goes through the region excluded by stellar cooling and NA64 invisible decay constraint.
There is still lot of parameter space left un-excluded  for $m_a\gtrsim 0.3$ MeV, $g_{ae}\lesssim 6\times 10^{-4}$, which can be probed by this boosted DM with $m_\chi< 1$ MeV. 
We restrict our analysis for DM masses down to $m_\chi\geq1$ keV, as lighter fermion DM is forbidden by the Pauli exclusion principle \cite{Tremaine:1979we}.
The blazar-boosted DM flux from BL Lacertae has less kinetic energy than TXS 0506+56, leading to weaker constraints coming from the former one. 

Finally, to draw the exclusion regions we assume $\langle\sigma v\rangle =0$ which is realistic for the case of asymmetric DM \cite{Petraki:2013wwa}.
However, the presence of a non-zero annihilation of DM to SM or other dark sector particles may deplete the spike density and thereby weaken the constraint.
For example, with $\langle\sigma v\rangle = 10^{-28}{\rm cm}^3{\rm s}^{-1}$ and $m_\chi=1$ MeV, the LOS density drops by factor $\sim 10^3$ (see Fig.\ref{den2}) which will weaken the constraint by factor $\sim 6$ since the rate depends on $\sim g_{ae}^4$.
In such case new regions probed by  BBDM become small.
For our minimal scenario DM can annihilate to only $e,a$ for $m_\chi> m_e,m_a$.
In that case, the constraints for heavier DM masses ($m_\chi \gtrsim 1$ MeV) are weaken due to less density. 
However, the constraints for $m_\chi\lesssim 10^{-1}$MeV $<m_a$  do not change at all for realistic cross-sections even in presence of a non-zero annihilation and thus our bounds are robust in this sense. 

\subsection{Vector portal}
Similarly, one can consider a vector portal DM where the mediator couples to both DM and $e$ like,
\begin{equation}
    \mathcal{L}\supset  g_{V\chi} V_\mu \bar{\chi} \gamma^\mu \chi + g_{Ve} V_\mu \bar{e} \gamma^\mu e.
    \label{eq:lag_vec}
\end{equation}
 $V$ is the vector mediator with mass $m_V$ and $g_{V\chi},g_{Ve}$ are the corresponding coupling strength with $\chi,e$.
There are several well motivated gauge extensions featuring such behavior \cite{Okada:2020cue}. 
As an example of UV completion of such scenario one may choose the  well studied $L_e-L_\tau$ model, where apart from $e,\tau,\nu_{e,\tau}$ \cite{Coloma:2022umy}, DM $\chi$ also couples to the BSM gauge boson $V$ \cite{Okada:2020cue}.
The differential cross-section of DM with high energy jet electron is given by,
%
%\begin{widetext}
\begin{eqnarray}
    \frac{d\sigma_{\chi e}}{dT_\chi}&=& \frac{1}{4\pi} \frac{g_{V \chi}^2 g_{V e}^2 }{(2 m_e T_e+T_e^2)(m_V^2+2 m_\chi T_\chi)^2} 
    \nonumber\\
    && \times \big[2 m_\chi (m_e+T_e)^2 -T_\chi \{(m_e+m_\chi)^2
    \nonumber\\
    &&+2 m_\chi T_e\}+m_\chi T_\chi^2 )\big],
    \label{eq:boost_v}
\end{eqnarray}
And the differential recoil cross-section is given by,
\begin{eqnarray}
    \frac{d\sigma_{\chi e}}{dE_R}&=& \frac{1}{4\pi} \frac{g_{V \chi}^2 g_{V e}^2 }{(2 m_\chi T_\chi+T_\chi^2)(m_V^2+2 m_e E_R)^2} \nonumber\\
    &&\times \big[2 m_e (m_\chi+T_\chi)^2 -E_R \{(m_i+m_\chi)^2
    \nonumber\\
    && +2 m_e T_\chi \}+m_e E_R^2 \big].
    \label{eq:recoil_v}
\end{eqnarray}

%\end{widetext}

\begin{figure}[!tbh]
    \centering
    \subfigure[\label{v1}]{
    \includegraphics[scale=0.45]{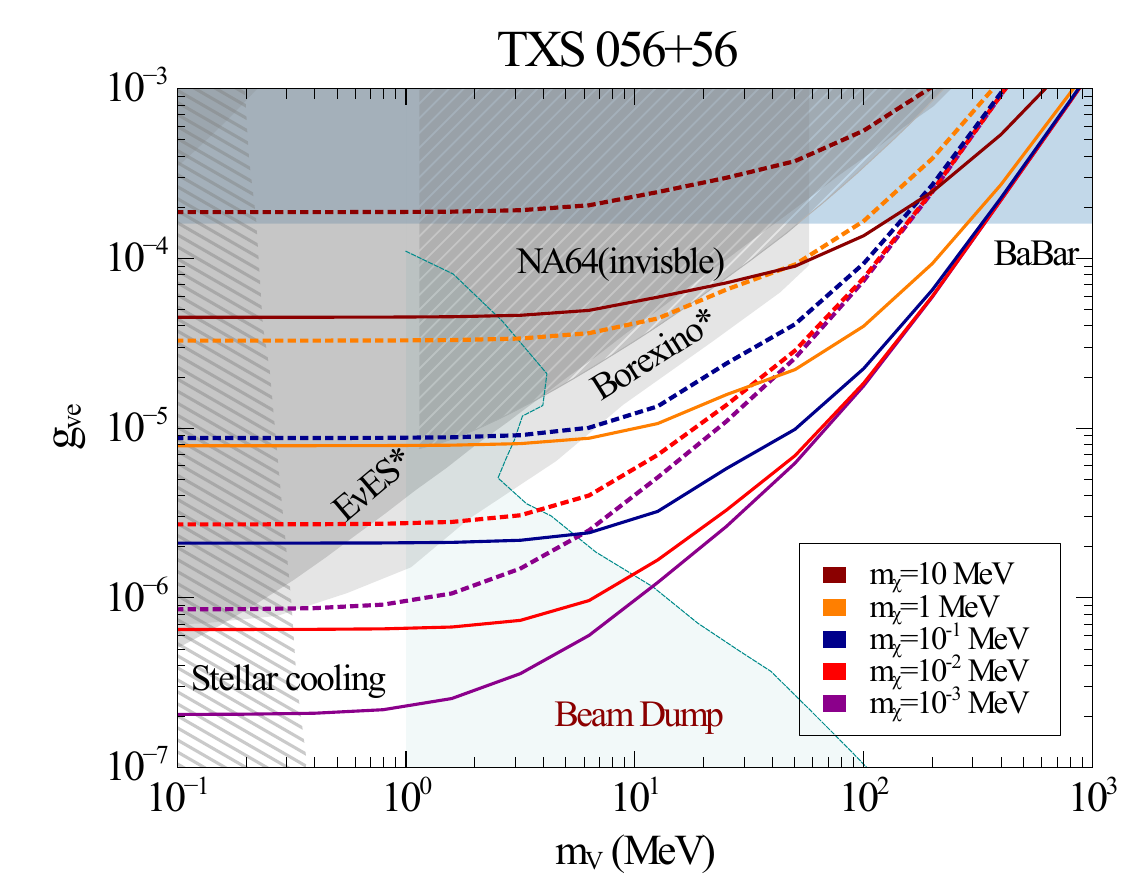}}
    \subfigure[\label{v2}]{
    \includegraphics[scale=0.45]{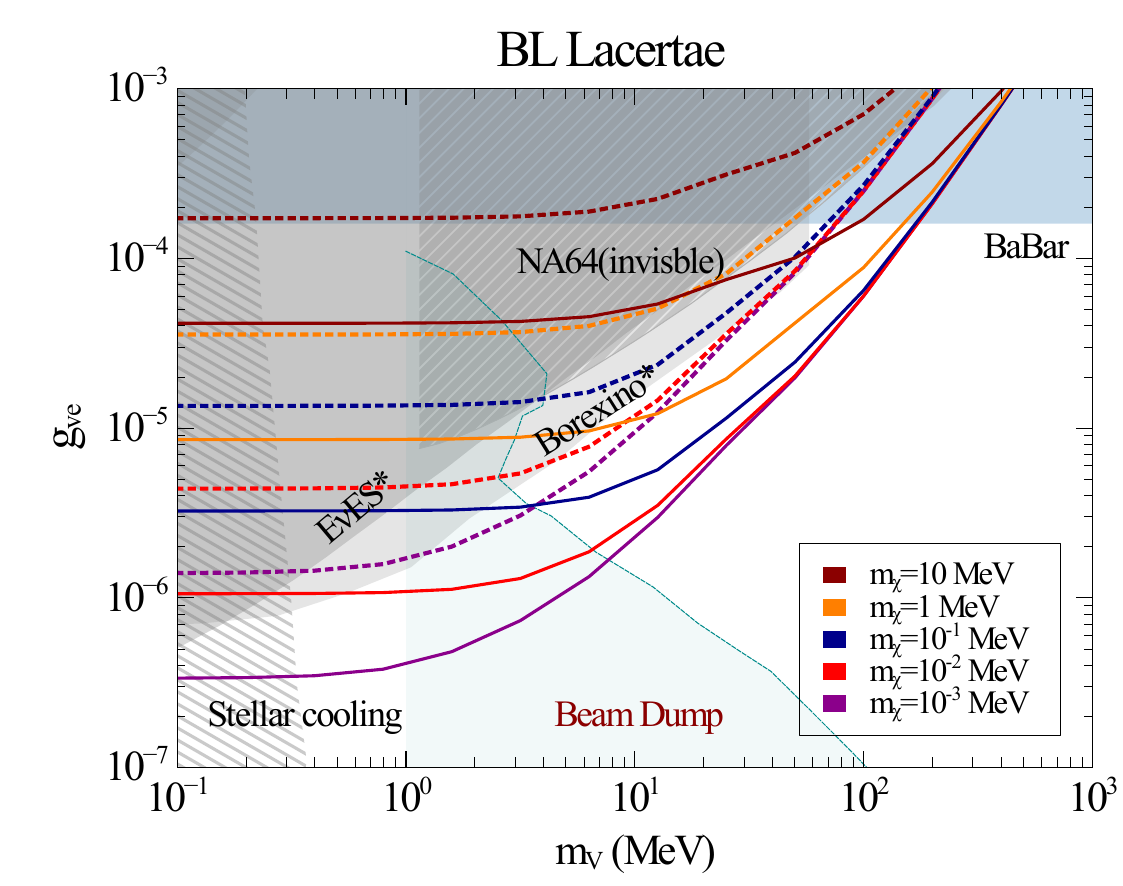}}
    \caption{$95\%$ CL constraints on BBDM in $m_V$ vs. $g_{_Ve}$ parameter space along with other existing constraints on electrophilic mediator considering (a) TXS 056+56 and (b) BL Lacertae as blazar source.  Regions above the individual colored lines are excluded from DM searches for different DM mass $m_\chi=$10 MeV, 1 MeV, 0.1 MeV, $10^{-2}$ MeV, $10^{-3}$ MeV denoted by  brown, orange, blue, red and magenta colored solid (dashed) lines for $\gamma=7/3$ ($\gamma=3/2$) respectively. The different shaded regions correspond to existing constraints on ALP discussed in the text. The constraints  with ``star" in label signify that they are applicable when the mediator couples to neutrino as well e.g. in $L_e-L_\tau$ model.}
    \label{fig:le_constraints}
\end{figure}

Following the same approach as in the previous subsection we evaluate the event rate at Super-K  and display the constraints on BBDM  in Fig.\ref{fig:le_constraints}. 
We show the constraints in the $m_V$ vs. $g_{_{Ve}}$ plane assuming TXS 0506$+56$ (BL Lacertae) as blazar source in Fig.\ref{v1} (Fig.\ref{v2}).
%Though we do not show the bounds assuming BL Lacertae as blazar source, from our observation in earlier subsection we expect qualitatively similar or slightly weaker constraints from it.
We show the limits on BBDM assuming $\gamma=7/3$ ($\gamma=3/2$) with solid (dashed) lines for different DM mass $m_\chi=$10 MeV, 1 MeV, 0.1 MeV, $10^{-2}$ MeV, $10^{-3}$ MeV following the same color convention as in the earlier sub-section. Here we also consider $\langle\sigma v\rangle =0$ to obtain the limits.

In the same plane (Fig.\ref{fig:le_constraints}) we showcase other existing constraints on electrophilic vector mediators. As shown earlier,  Babar ($g_{_{Ve}}\gtrsim10^{-4}$) \cite{BaBar:2014zli,Bauer:2018onh} and NA64 ``invisible" \cite{Banerjee:2019pds} place strongest constraints on couplings with higher strength.
On the other hand, for lower values of the coupling, bounds from beam dump experiments \cite{Bauer:2018onh} are
applicable. Constraints from stellar cooling are applicable  for $m_V\lesssim 0.4$ MeV \cite{Hardy:2016kme}.
Even though we do not mention explicitly in the plot, the invisible decay constraints (BaBar, NA64) are relevant as long as $m_a> 2 m_\chi$ and visible decay constraints (Beam Dump) are applicable for only $m_a<2 m_\chi $ as elaborated earlier in detail.
If the vector couples to neutrinos e.g. in models like $L_e-L_\tau$ \cite{Ghosh:2024cxi}, then neutrino scattering constraints from E$\nu$ES from XENON \cite{Majumdar:2021vdw,DeRomeri:2024dbv}, Borexino \cite{Coloma:2022umy,Khan:2019jvr,AtzoriCorona:2022moj} provide the strongest constraint.
Despite all the stringent constraints, the limits set by BBDM search exclude a significant portion of new parameter space for $m_\chi\lesssim 10$ MeV.
These constraints are significantly stronger than the cosmic ray boosted DM search \cite{Guha:2024mjr}.
Although the effective vector portal (without $\nu-V$ couplings) blazar boosted DM at heavy mediator limit has been investigated earlier in ref.\cite{Bardhan:2022bdg}, here we obtain the bounds in the whole mass vs. coupling plane of the mediator.
In the absence of neutrino coupling of the mediator, BBDM explores new parameter space in the light mediator regime too as shown in Fig.\ref{fig:le_constraints}. 
Note that for vector mediator there is no momentum suppression in the cross-section which makes the limit on the coupling almost $\mathcal{O}(1)$ order (cross-section $\sim \mathcal{O}(10^4)$ ) stronger than ALP mediated scenario. 
The bounds assuming BL Lacertae as blazar source are weaker than with TXS 56+056 as expected from our discussion  in earlier subsection (see the discussion of Fig.\ref{fig:constraints}).
Before concluding the discussion we highlight that the cosmological constraint on relativistic degrees of freedom $N_{\rm eff}$ excludes electrophilic mediators with mass $\lesssim \mathcal{O}(1)$ MeV and coupling $\gtrsim 10^{-9}$ \cite{Ghosh:2024cxi,Ibe:2019gpv}. Our exclusion limits probe new parameter space despite of these constraints too. $N_{\rm eff}$ also excludes thermal DM for mass $\lesssim 3$ MeV \cite{Escudero:2018mvt}. However, such constraints are dependent on the underlying UV complete DM models, and can be subjugated by considering an extended dark sector.
Thus blazar boosted scenario not only sets the strongest direct search limit in the (sub) MeV DM masses, rather it also probes previously unconstrained parameter space for different mediators contrary to other boosted DM scenarios.

\section{DM with both nucleon and electron coupling}
\label{sec:nucleo}
%========================
So far we have discussed about DM that is boosted by only blazar electron jet. However,  there exist several well motivated mediators that couple to both electron and proton (quarks), so that DM can be potentially up-scattered by both electrons and protons in the blazar jet.
The minimal and maximal boost factor of the proton jet in the blob frame given by SED modeling are 
$\gamma'_{\rm min, p}=1.0$ and $\gamma'_{\rm max, p}=5.5 \times 10^7$ ($\sim 10^4$ times larger than electrons) for TX056$+56$ \cite{Granelli:2022ysi}.
The gain in DM being up-scattered by proton as well is that boosted DM can now have higher flux even at energy $\sim 10^3$ GeV contrary to the electrophilic case \cite{Granelli:2022ysi}.
Thus considering both the effects of proton and electron may lead to higher flux of BBDM resulting in more stringent bounds. To obtain the blazar proton jet of TX056$+56$  we use the same eq.\eqref{eq:spectrum} with the obvious replacements ($i\equiv p$): $m_e\to m_p$, $\alpha_e\to\alpha_p(=2)$, and $L_e\to L_p(=2.55 \times 10^{48}$ erg/s) \cite{Granelli:2022ysi}. The normalization constant $c_e$ is also replaced by $c_p$ normalized with respect to $L_p$. The values of these parameters corresponding to BL Lacertae are $\gamma'_{\rm min, p}=1.0$ and $\gamma'_{\rm max, p}=1.9 \times 10^9$, $\alpha_p=2.4$  and $L_p=9.8 \times 10^{48}$ erg/s \cite{Granelli:2022ysi}.

\subsection{ALP portal}
For ALP portal DM scenario we consider the Yukawa like couplings of ALP with quarks and electrons.
\begin{equation}
    \mathcal{L}\supset \sum_{f=q,e}i \frac{c}{f_a}m_f a \bar{f} \gamma^5 f,
\end{equation}
where, the sum is over quarks and electron. This type of Lagrangian is motivated from DFSZ models \cite{Bharucha:2022lty}.
For calculating DM up-scattering by electrons  we use the same eq.\eqref{eq:boost_a} with $ g_{ae}=\frac{c}{f_a}m_f$. 
While for the case of proton up-scattering we replace $g_{ae}$ by $g_{ap}\approx 0.45 c/f_a m_u (m_p/(m_u+m_d))$ and multiply the whole equation with the square of a form factor given by, $F_a(q^2)=F_{\rm axial}(q^2) C_q/(q^2+M_q^2)$, with $C_q=0.9,M_q=0.33$ GeV, and $F_{\rm axial}(q^2)=1/(1+q^2/\Lambda_a^2)^2~(\Lambda_a=1.32$ GeV) \cite{10.21468/SciPostPhys.10.3.072}.
To evaluate the incoming BBDM flux we add up both the contributions from electron and proton up-scattering \cite{Granelli:2022ysi}.
While computing the recoil rate, we consider only electron scattering for conservative estimations since, with TeV energy and feeble couplings, DM-proton scattering will be suppressed in the detector.

\begin{figure}[!tbh]
    \subfigure[\label{cf1}]{
    \includegraphics[scale=0.45]{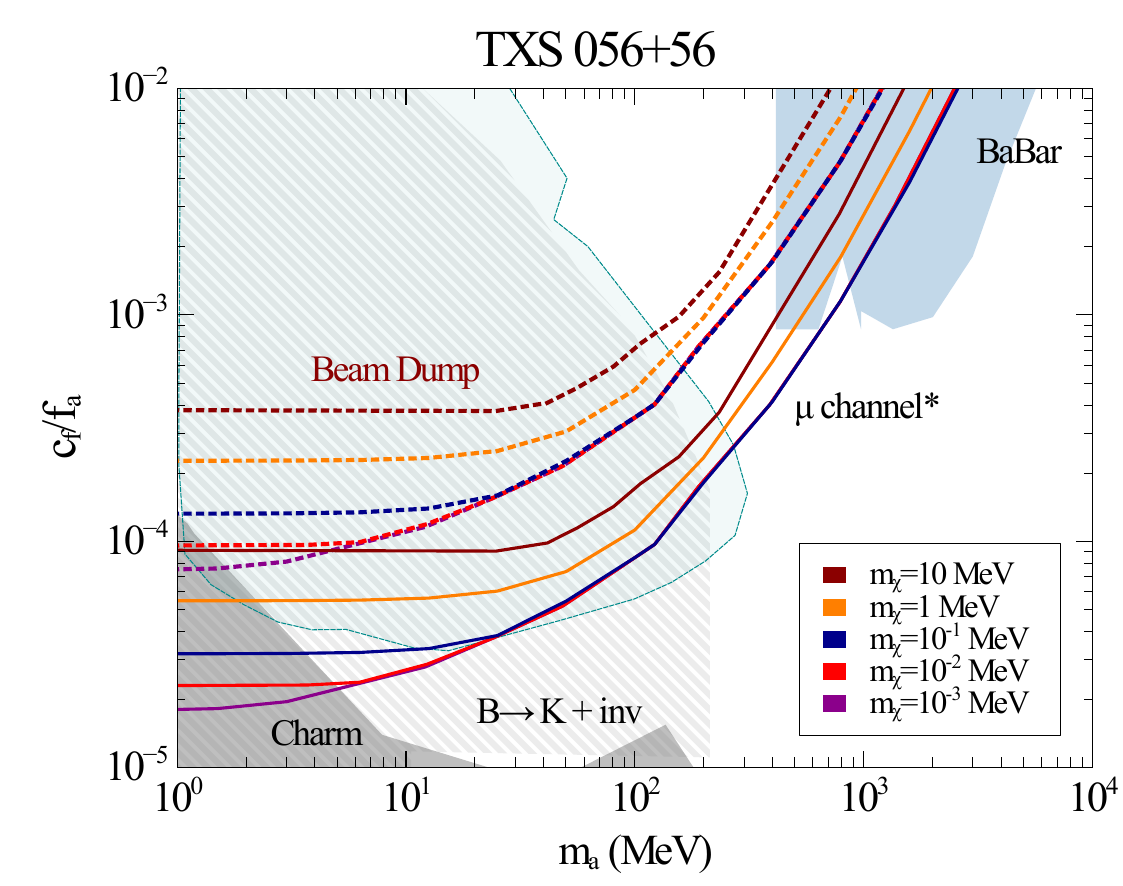}}
    \subfigure[\label{cf2}]{
    \includegraphics[scale=0.45]{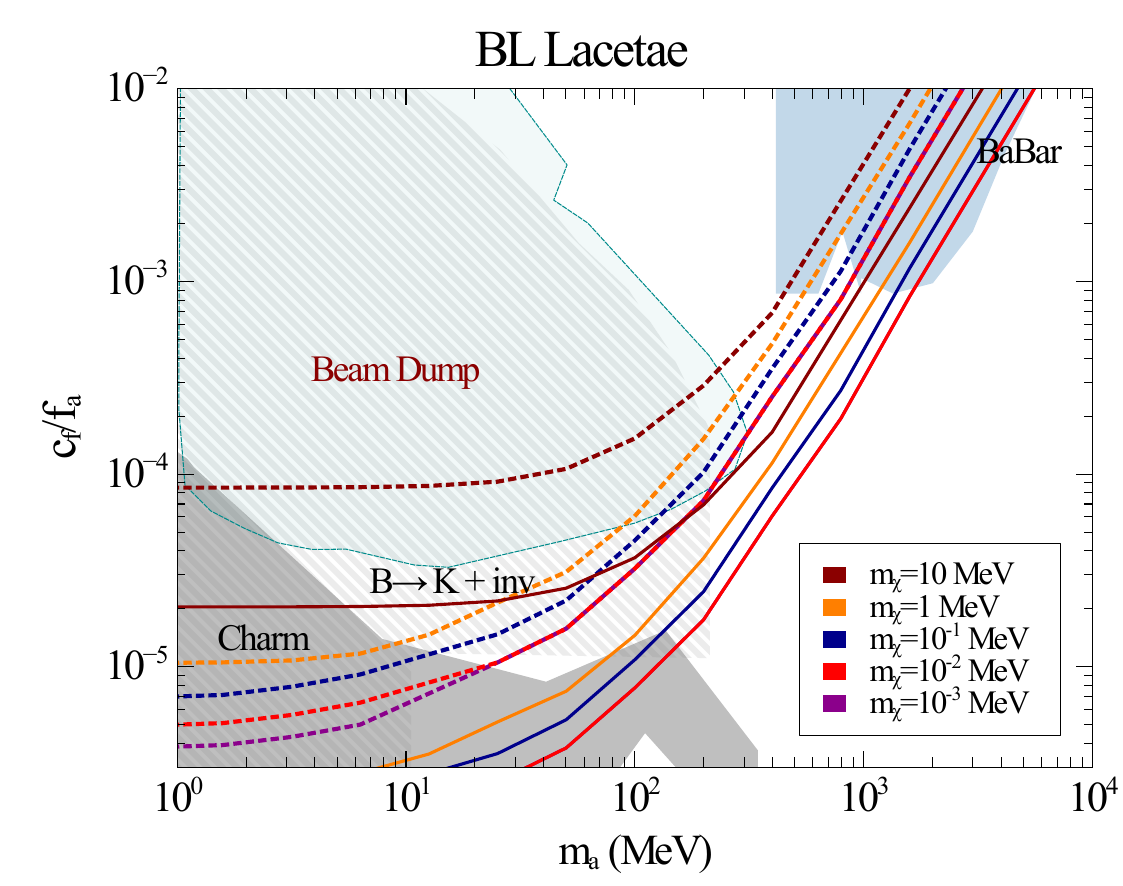}}
    \caption{$95\%$ CL constraints on BBDM in $m_a$ vs. $c_f/f_{a}$ parameter space along with other existing constraints on ALP considering (a) TXS 056+56 and (b) BL Lacertae as blazar source.  {Regions above} the individual colored lines are excluded from DM searches for different DM mass $m_\chi=$10 MeV, 1 MeV, 0.1 MeV, $10^{-2}$ MeV, $10^{-3}$ MeV denoted by  brown, orange, blue, red and magenta colored solid (dashed) lines for $\gamma=7/3$ ($\gamma=3/2$) respectively. The different shaded regions correspond to existing constraints on ALP discussed in the text. }
    %The thin lines indicate constraints applicable in presence of ALP muon coupling.
    \label{fig:alp_constraints}
\end{figure}

We show the constraints on $m_a$ vs. $c/f_a$ plane in Fig. \ref{fig:alp_constraints} and to signify the limits we follow the same convention as earlier used in this paper. 
The limits obtained assuming TXS 0506$+56$ (BL Lacertae) as blazar source are displayed in Fig.\ref{cf1} (Fig.\ref{cf2}).
Note that, the mass suppression in the aforementioned Yukawa like interaction makes the limits look apparently weaker than the pure elctrophilic case.
One may recast the bound on  $g_{ae}(=m_e c/f_a)$ as well and perceive that limits on the effective couplings are way stronger than the electrophilic case (\ref{fig:alp_constraints}).
Due to the presence of quark couplings strong constraints come from meson decay (invisible) in the low $m_a$ regime shown by the gray shaded region (Charm) and light gray diagonal shaded region ($B\to K+$invisible) \cite{Dolan:2014ska}. In the high mass regime there exists constraint from BaBar \cite{Dolan:2014ska}.
Recall that these are invisible decay constraints and applicable for only $m_a> 2 m_\chi$. On the other hand the beam dump constraints \cite{Bjorken:1988as} are relevant for only $m_a<2 m_\chi$.
For the same reasons kaon decay (to $e^+e^-/\mu^+\mu^-$) constraints are not applicable for our concerned parameter space \cite{Bharucha:2022lty}. 
%The contour with no shade signify the constraint from meson decay to di-muon channel \cite{Bharucha:2022lty}. However, in the absence of ALP-$\mu$ coupling this constraint is not applicable. 
Despite all these stringent constraints, there still remains some open parameter space at $m_a\sim200-400$ MeV which is constrained by BBDM searches assuming our minimal scenario.
Note that, in contrast to electrophilic DM (sec.\ref{sec:electro}), here the limits obtained with BL Lacertae as blazar source are slightly stronger than the ones with  TXS 056+56.
Since  the value of $\gamma'_{\rm max, p}$  for BL Lacertae is $\sim 10^2$ higher than TXS 056+56, the corresponding proton jet of BL Lacertae has higher kinetic energy than that of TXS 056+56. This results in DM upscattered with higher kinetic energy, leading to an enhanced recoil rate in Super-k while considering BL Lacertae as blazar source \cite{Granelli:2022ysi}.

\begin{figure}[!tbh]
    \includegraphics[scale=0.45]{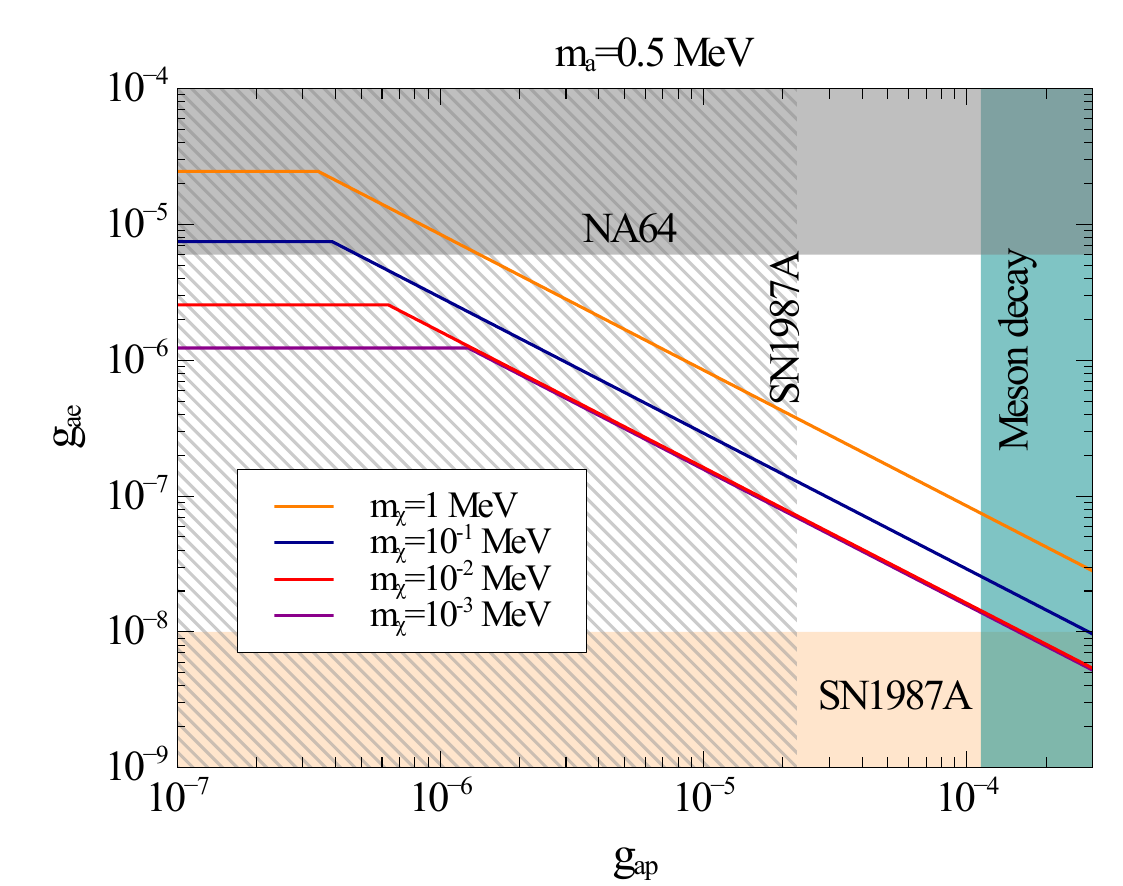}
    \caption{$95\%$ CL constraints on BBDM in $g_{ap}$ vs. $g_{ae}$ plane for fixed $m_a=0.5$ MeV for $\gamma=7/3$ considering TXS 056$+56$ as blazar source. Shaded regions correspond to the existing constraints on the couplings. Limits corresponding to $m_\chi\in 10^{-3},10^{-2},10^{-1},1$ MeV are shown by magenta, red, blue and orange colors respectively.}
    \label{fig:gae_gap}
\end{figure}

To show the constraints in terms of electron and proton couplings, one may also assume effective ALP couplings like $\mathcal{L}\sim i g_{ae} a\bar{e}\gamma^5 e+ i g_{ap} a\bar{p}\gamma^5 p$ being agnostic about the origin.
The constraints on BBDM indeed explore some new parameter space in $g_{ap}$ vs. $g_{ae}$ plane for fixed $m_a$ as shown in Fig.\ref{fig:gae_gap} for $\gamma=7/3$.
We show the result for a fixed ALP mass $m_a=0.5$ MeV along with the existing constraints on the couplings \cite{Bell:2023sdq,Waites:2022tov}. While considering effective proton coupling of ALP {there is a window} of unconstrained region around $0.4~{\rm MeV}\lesssim m_a\lesssim 1$ MeV \cite{Bell:2023sdq} which also remains un-excluded for electron coupling.
The direct search constraints on BBDM is shown in the same plane with different lines for $m_\chi\in 10^{-3},10^{-2},10^{-1},1$ MeV shown by magenta, red, blue and orange color respectively. The regions above the contour lines are excluded from Super-K.
For $g_{ap}\gtrsim 10^{-4}$ the region is excluded by meson invisible decay (for $m_a\leq 0.1$ MeV) and $g_{ap}\lesssim 2 \times 10^{-5}$ is excluded by supernova cooling \cite{Bell:2023sdq}.
NA64 visible decay constraint on $g_{ae}<2 \times 10^{-6}$ is applicable to only $m_\chi=1$ MeV.
For smaller values of $g_{ap} (\lesssim10^{-6})$, the constraints are independent of $g_{ap}$ and hence the limits become flat.
{If one considers higher values of ALP-proton coupling ($g_{ap}\gtrsim 10^{-6}$) the constraints on $g_{ae}$ become stronger and the limits also exclude previously unexplored parameter space.} 
%Considering higher values of proton coupling  indeed sets stronger limit on $g_{ae}$ as well which explore previously un-excluded parameter space.

\subsection{Vector portal}
%========================================
Similarly, one may consider a vector like mediator which couples to both nucleon and {an electron} as in $B-L$ gauge extension \cite{AtzoriCorona:2022moj}.
In such scenario the gauge boson couples as 
\begin{equation}
    \mathcal{L}\supset \sum_{f=q,e} g_{B-L} q_f  \bar{f} \gamma^\mu f V_\mu,
\end{equation}
where, the sum is over quarks and electron. $q_f$ is the $B-L$ gauge charge and $q_f=-1~(1/3)$ for leptons (quarks). The coupling with DM is same as in eq.\eqref{eq:lag_vec}.
For calculating DM up-scattering by electrons  we use the same eq.\eqref{eq:boost_v} with $ g_{ae}=g_{B-L}$, while for the case of proton we replace $g_{ap}=g_{B-L}$ and multiply the whole equation with the square of a form factor given by, $F_V(q^2)=(1+q^2/\Lambda_V^2)^{-2}$, with $\Lambda_V=0.77$ GeV \cite{Guha:2024mjr}.
Analogous to the previous section we add up the incoming BBDM  flux due to both proton and electron up-scattering.

\begin{figure}[!tbh]
    \centering
    \subfigure[\label{bl1}]{
    \includegraphics[scale=0.45]{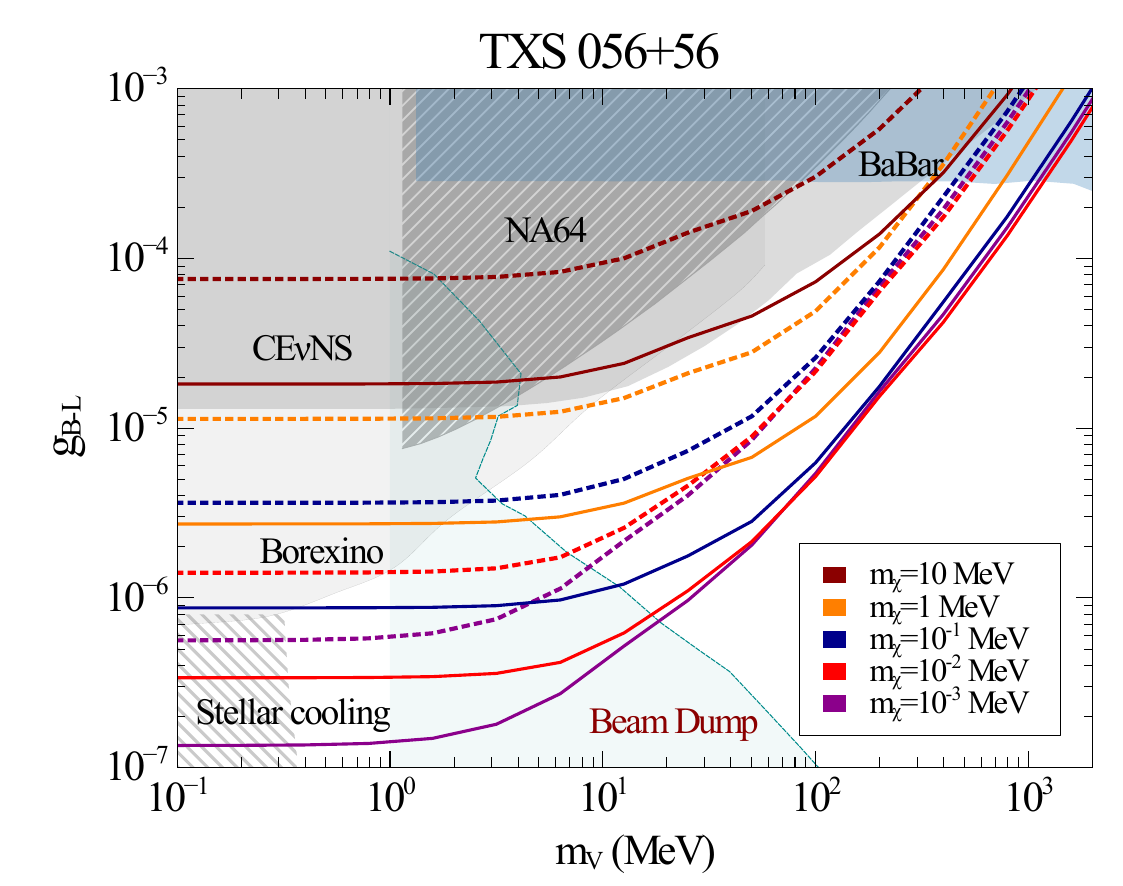}}
    \subfigure[\label{bl2}]{
    \includegraphics[scale=0.45]{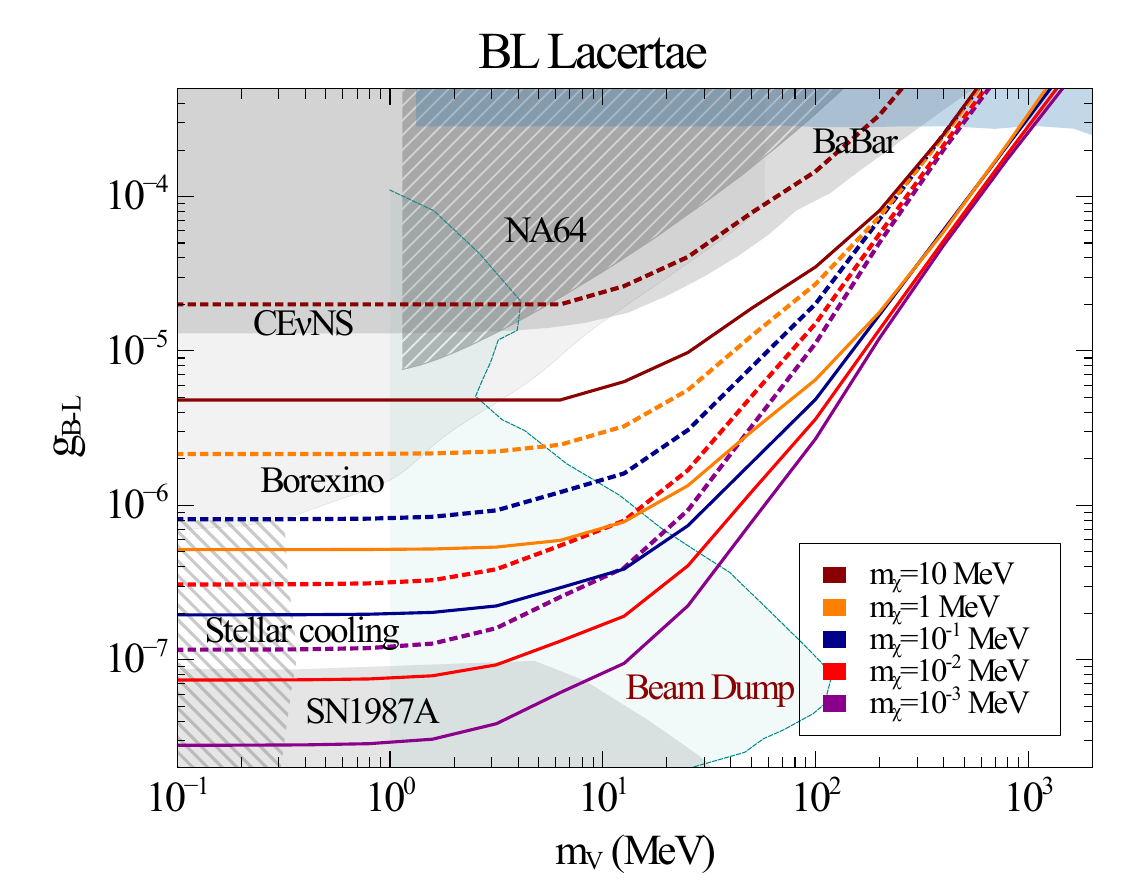}}
    \caption{$95\%$ CL constraints on BBDM in $m_V$ vs. $g_{B-L}$ parameter space along with other existing constraints on $B-L$ gauge mediator considering (a) TXS 056+56 and (b) BL Lacertae as blazar source.  Regions above the individual colored lines are excluded from DM searches for different DM mass $m_\chi=$10 MeV, 1 MeV, 0.1 MeV, $10^{-2}$ MeV, $10^{-3}$ MeV denoted by  brown, orange, blue, red and magenta colored solid (dashed) lines for $\gamma=7/3$ ($\gamma=3/2$) respectively. The different shaded regions correspond to existing constraints on ALP discussed in the text.} %The constraints  with ``star" in label signify that they are applicable when the mediator couples to neutrino as well e.g. in $L_e-L_\mu$ model.
    \label{fig:constraints_bl}
\end{figure}

{Finally we place the $95\%$ exclusion limit from Super-K in the $m_V$ vs. $g_{B-L}$ plane as shown in Fig.\ref{fig:constraints_bl}.
The constraints corresponding to different DM masses are depicted by the same color convention as used earlier in Fig.\ref{fig:constraints}.}
Limits obtained assuming TXS 0506$+56$ (BL Lacertae) as blazar source are displayed in Fig.\ref{bl1} (Fig.\ref{bl2}).
In the same plane we display the existing constraints from beam dump experiments \cite{Bauer:2018onh}, neutrino scattering constraints from E$\nu$ES and CE$\nu$NS in XENON \cite{Majumdar:2021vdw,DeRomeri:2024dbv}, Borexino \cite{Coloma:2022umy,Khan:2019jvr,AtzoriCorona:2022moj} and stellar cooling \cite{Hardy:2016kme}.
Despite all the stringent constraints the limits set by BBDM search exclude significant portion of new parameter space for $m_\chi\lesssim 10$ MeV.
Since the DM flux is higher due to proton scattering compared to the electrophilic case, in the heavy mediator limit the constraints on the coupling $g_{B-L}$  are almost $4$ times stronger than in Fig.\ref{fig:le_constraints} and hence the bounds on the cross-sections ($\sim g_{B-L}^2$) will be almost $\gtrsim\mathcal{O}(1)$ stronger.
In particular, the bounds with BL Lacertae on $g_{B-L}$ (Fig.\ref{bl2}) is stronger than $g_{Ve}$ (Fig.\ref{v2}) by$\sim\mathcal{O}(1)$ factor due to the higher kinetic energy of associated blazar proton jet. 
Thus combining both the effect of electrons and protons from blazar source can lead to stronger limits on MeV DM beyond the existing constraints on the mediator.

\section{Conclusion}
\label{sec:conc}
%======================= 
Multi-ton direct detection experiments face severe challenge in detecting low mass galactic DM even with cutting-edge instruments and large detector volume. 
To circumvent this issue one may adopt the boosted DM scenario which can potentially exclude the previously unexplored DM mass region.
Generally significant portion of the exclusion limits obtained in such scenario {may lie within already existing} constraints {once we consider BSM mediators} as a portal between DM and SM. 
The situation is even more disappointing for pseudoscalar mediators where the momentum suppression leads to very weak constraints on the cross-section.
{Such exclusion limits fail to explore any new parameter space and almost touch the unitarity limit in the mass-coupling parameter space of the mediator \cite{Bell:2023sdq}.
This issue can be disentangled by considering low mass DM boosted by blazar jet  which serves as the main goal of this work.}
We embark on a minimal BSM model, where a DM $\chi$ couples to SM via a pseudoscalar mediator, the so called ``ALP portal DM". We estimate the expected recoil rate in Super Kamiokande triggered by such blazar boosted DM for different model parameters.
For comparison we consider two benchmark blazar sources: TXS 0506$+56$ and a less distant BL Lacertae. 
The spike density of DM and blazar electron jet (with maximum energy $\sim 10^4$ GeV) leads to a significant amount of BBDM flux reaching at Earth despite their large distances.
We obtain $95\%$ CL upper limit on DM-$e$ interaction in the mass vs. coupling parameter space of the mediator ALP. 
{From our analysis, we summarize our major findings below:
\begin{itemize}
    \item Stringent constraints on the ALP parameter space arise from astrophysics and ground based searches like BaBar, NA64 and beamp dump experiments, yet our limits exclude significant new parameter space ($g_{ae}\gtrsim 10^{-4} ~(2 \times 10^{-5})$ for $ m_\chi= 1$ MeV (0.1 MeV) with $m_a/m_\chi=10$) as shown in Fig.\ref{fig:constraints}.
   \item For completeness we also consider a vector portal DM model and following the same approach we obtain the exclusion limits as shown in Fig.\ref{fig:le_constraints}. The BBDM search sets  stronger limits on the mass vs. coupling plane ($g_{aV}\gtrsim 10^{-5} ~(2 \times 10^{-6})$ for $ m_\chi= 1$ MeV (0.1 MeV) with $m_a/m_\chi=10$) compared to other BDM scenarios investigated in the literature \cite{Bell:2023sdq,Guha:2024mjr}.
\item Finally we allow the mediators to couple with both electron and nucleon for both ALP and vector portal DM scenario.
Thanks to the higher kinetic energy (maximum energy $\sim 10^7$ GeV) of blazar protons, these type of interactions lead to further improvement in the sensitivity of BBDM searches at Super-K \cite{Granelli:2022ysi}.
We find that even smaller values of BSM couplings  responsible for the interaction between low mass DM and SM can be probed  combining the effect of multiple sources for DM up-scattering (i.e. both blazar electron and proton jets) as shown in Fig.\ref{fig:alp_constraints} (ALP portal) and Fig.\ref{fig:constraints_bl} (vector portal).
Such characteristics are quite generic for models featuring a mediator with universal couplings to SM sector \cite{Bharucha:2022lty,Okada:2018ktp}.
\end{itemize}
}

Thus one can explore the uncharted territories of light DM phenomenology with blazar boosted scenario considering concrete UV complete (low mass) DM models. 
Upon successful identification of proper background, experiments like IceCube and future experiments like DUNE, JUNO have the potential to probe even 
smaller interaction strength between the dark and visible sectors. 
{It would be also interesting to explore the effects of different models of various blazar sources e.g. ref.\cite{Potter:2015tya,refId0} in the context of light DM searches.}

%\section{Possibility}
%Both $\nu$ and $e$

\section*{Acknowledgment}
%%%%%%%%%%%%%%%%%%%%%%%%%%%
We would like to thank Dilip K. Ghosh and Filippo Sala for the careful reading of the manuscript and the helpful comments. We also thank Rohan Pramanick, Pratick Sarkar for fruitful discussions and Gonzalo Herrera for the useful email conversations.
It is our pleasure to thank Vikramaditya Mondal for providing computational support. 
We acknowledge the local hospitality at ICTS Bengaluru during the visit for GWBSM 76 program.
This work is funded by CSIR, Government of India under the NET JRF fellowship scheme with
Award file No. 09/080(1172)/2020-EMR-I.

\appendix
\section{Energy spectrum of blazar}
\label{sec:energy}
\counterwithin{figure}{section}
%==========================
%We assume the blob is moving in the $z$ axis from the BH
%rest frame with velocity $\beta_B$. The corresponding Lorentz factor of the blob is given by $\Gamma_B=1/\sqrt{1-\beta_B^2}$.
Following our discussion earlier in Sec.\ref{sec:spec} we consider the particles are going outwards in the blob frame with velocity $\beta'$ and an angle $\theta'$ with respect to the $z$ axis.
%The prime quantities are defined in blob frame.
The velocities of the jet particles in blob frame in parallel direction and perpendicular direction are given by proper Lorentz transformation,
\begin{eqnarray}
    \beta'_{\|}= \dfrac{\beta \cos\theta -\beta_B}{1-\beta_B \beta \cos\theta} ,~~~ \beta'_{\perp}=\dfrac{\beta \sin\theta}{\Gamma_B(1-\beta_B \beta \cos\theta)}.
    \label{eq:beta}
\end{eqnarray}
Thus one can obtain the total velocity $\beta'$ in terms of the two components and
the boost factor $\gamma'=E'/m =(1- {\beta'}^2)^{-1/2}$ in the blob frame can be found as,
\begin{eqnarray}
        \gamma'&=&\gamma \Gamma_B(1-\beta_B \beta \cos\theta), 
     \label{eq:gam}
\end{eqnarray}
%==========================
where, $\gamma=E/m$ is the boost factor in the observer frame. Another important quantity required to estimate the flux is the polar angle $\theta$ (in observer frame), which can be related to the polar angle $\theta'$ (in blob frame by) as,
\begin{eqnarray}
    \cos{\theta'}&=& \dfrac{(\beta \cos\theta -\beta_B)}{\left[(1-\beta \beta_B \cos{\theta})^2 -(1-\beta^2) (1-\beta_B^2)\right]^{\frac{1}{2}} },~~
    \label{eq:mup}
\end{eqnarray}
using the same eq.\eqref{eq:beta}.

For the sake of simplicity we define $\mu= \cos{\theta},$ and $\mu'= \cos{\theta'}$.
The number of particle emitted from the blazar in per unit time / unit energy and  unit solid angle is defined in the blob frame as $d\Gamma'/dE' d\Omega'$. Analogously the spectrum in observer frame is defined as,
\begin{eqnarray}
    \frac{d\Gamma}{dE d\Omega} &=& \Gamma_B\frac{d\Gamma'}{dE' d\Omega'} \Big|\frac{\partial(E',\mu')}{\partial(E,\mu)} \Big|,
    \label{eq:det}
\end{eqnarray}
where the second term is the Jacobian of the transformation $J$ and can be obtained using eq.\eqref{eq:gam}-eq.\eqref{eq:mup}. {Plugging  $J$ in eq.\eqref{eq:det}} one finds,
\begin{eqnarray}
    \frac{d\Gamma}{dE d\Omega} &=& \frac{d\Gamma'}{dE' d\Omega'} \times \nonumber \\
    &&\dfrac{\beta \mu-\beta_B}{\sqrt{(1-\beta_B\beta \mu)^2-(1-\beta^2)(1-\beta_B^2)}}.
    \label{eq:trans}
\end{eqnarray}

\subsection*{Normalization:}
\label{sec:norm}
The normalization constant in eq.\eqref{eq:spectrum} is evaluated by normalizing the total luminosity in the observer's frame with the observed one $L_i$ (for particle $i$).
\begin{eqnarray}
    L_i&=& \int dE d\Omega \frac{d\Gamma_i}{dE_i d\Omega} E 
\end{eqnarray}
However, looking at the expression of the spectrum it is expected that the integration is easier in blob frame. So, using the Jacobians we recast the integrand in the blob frame as,
\begin{eqnarray*}
    L_i
    &=& \int dE' \frac{d\Gamma'_i}{dE'_i d\Omega'} E_i' d\Omega' \frac{1}{(1-\beta_B \beta \cos\theta)}.
\end{eqnarray*}
Since most of the emitted particles are relativistic we can assume $\beta\approx 1$ and inverting eq.\eqref{eq:mup} we get 
\begin{eqnarray*}
    \cos \theta= \frac{\cos{\theta'}+\beta_B}{1+\beta_B \cos{\theta'}}, ~{\rm and}~1-\beta_B \cos\theta= \frac{1-\beta_B^2}{1+\beta_B \cos{\theta'}}
\end{eqnarray*}
Substituting this in the above expression of $L_i$ we get,
\begin{eqnarray*}
     L_i&=& \int dE' \frac{d\Gamma'_e}{dE'_i d\Omega'} E_i' d\Omega' \frac{1+\beta_B \cos{\theta'}}{1-\beta_B^2} \\
     &=& 2\pi \frac{1}{1-\beta_B^2} 2 \frac{1}{4\pi} \int dE' \big(\frac{E'}{m_i}\big)^{-\alpha_i}E'\\
     &=& \Gamma_B^2 m_i^2 \int_{\gamma_{\rm min}}^{\gamma_{\rm max}} d\gamma' (\gamma')^{-\alpha_i} 
\end{eqnarray*}
Hence one can find the normalization constant,
\begin{eqnarray}
c_i&=& \frac{L_i}{\Gamma_B^2 m_i^2} \label{eq:ce} \\ 
 && \nonumber \times\Bigg \{ 
  \begin{array}{ c l }
    1/(\log (\gamma'^{\rm max}/\gamma'^{\rm min}) ) & \quad \textrm{if } \alpha_i=2 \\
   \dfrac{ (2-\alpha_i)}{ \big[ (\gamma'_{\rm max})^{2-\alpha_i} -(\gamma'_{\rm min})^{2-\alpha_i} \big] }              & \quad \textrm{if } \alpha_i\neq2
  \end{array}
\end{eqnarray}
\\

%%%%%%%%%%%%%%%%%%%%%%%%%%%%%%%5
\begin{figure}[!tbh]
    \centering
    \includegraphics[scale=0.4]{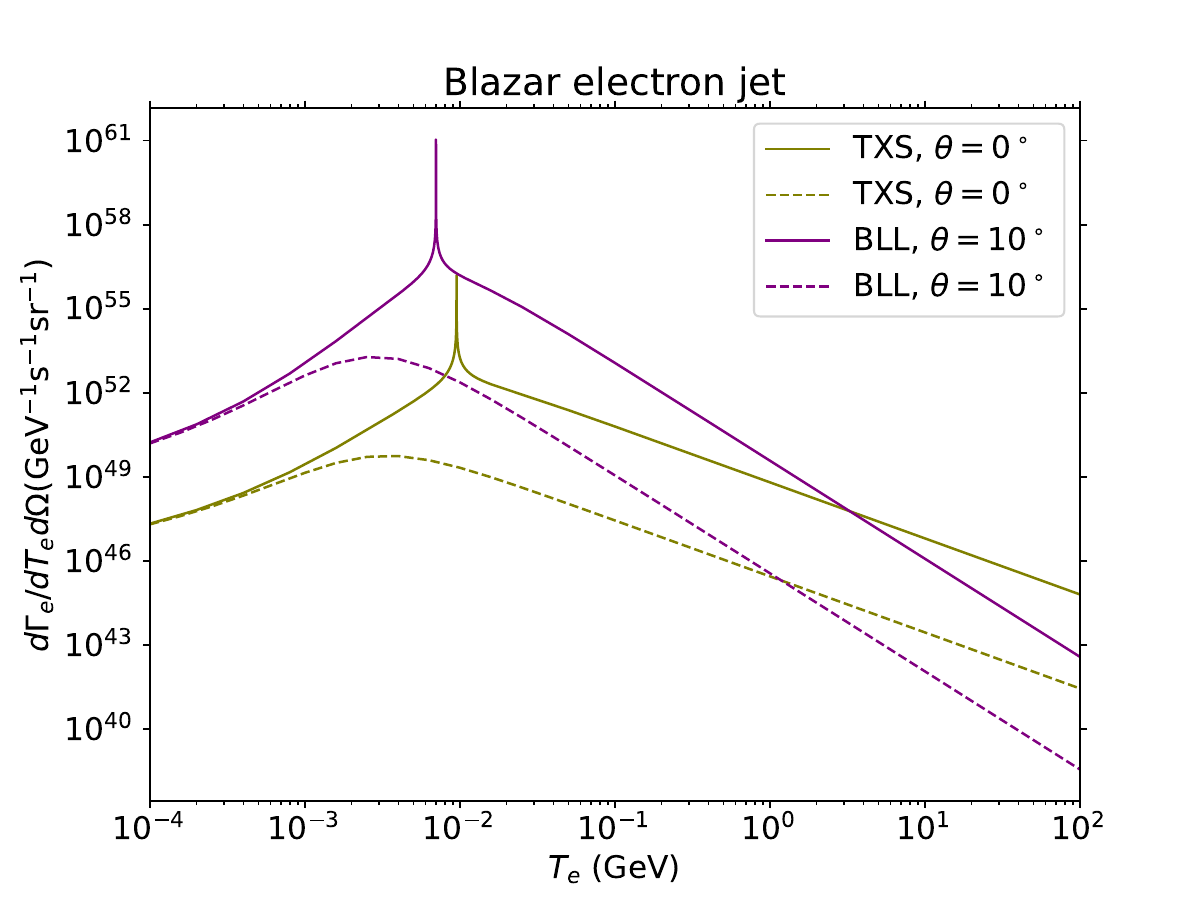}
    \caption{Blazar electron jet flux for TXS 0506$+56$ (BL Lacertae) are shown with olive (magenta) color for two different values of $\theta (\cos^{-1}\mu)=0^\circ, 10^\circ$ shown by solid and dashed lines respectively.}
    \label{fig:ejet}
\end{figure}

The corresponding blazar electron jet flux in observer's frame for the aforementioned two blazar source is shown in Fig.\ref{fig:ejet}.
The flux due to TXS 0506$+56$ (BL Lacertae) are shown with olive (magenta) color for two different values of $\theta (\cos^{-1}\mu)=0^\circ, 10^\circ$ depicted by solid and dashed lines respectively.

%&%%%%%%%%%%%%%%%%%%%%%%%%%%%%%%%%%%
\section{Spike density of DM}
\label{apx:dm_density}
%==============
%
\begin{figure}[!tbh]
    \centering
    \subfigure[\label{den1}]{
    \includegraphics[scale=0.4]{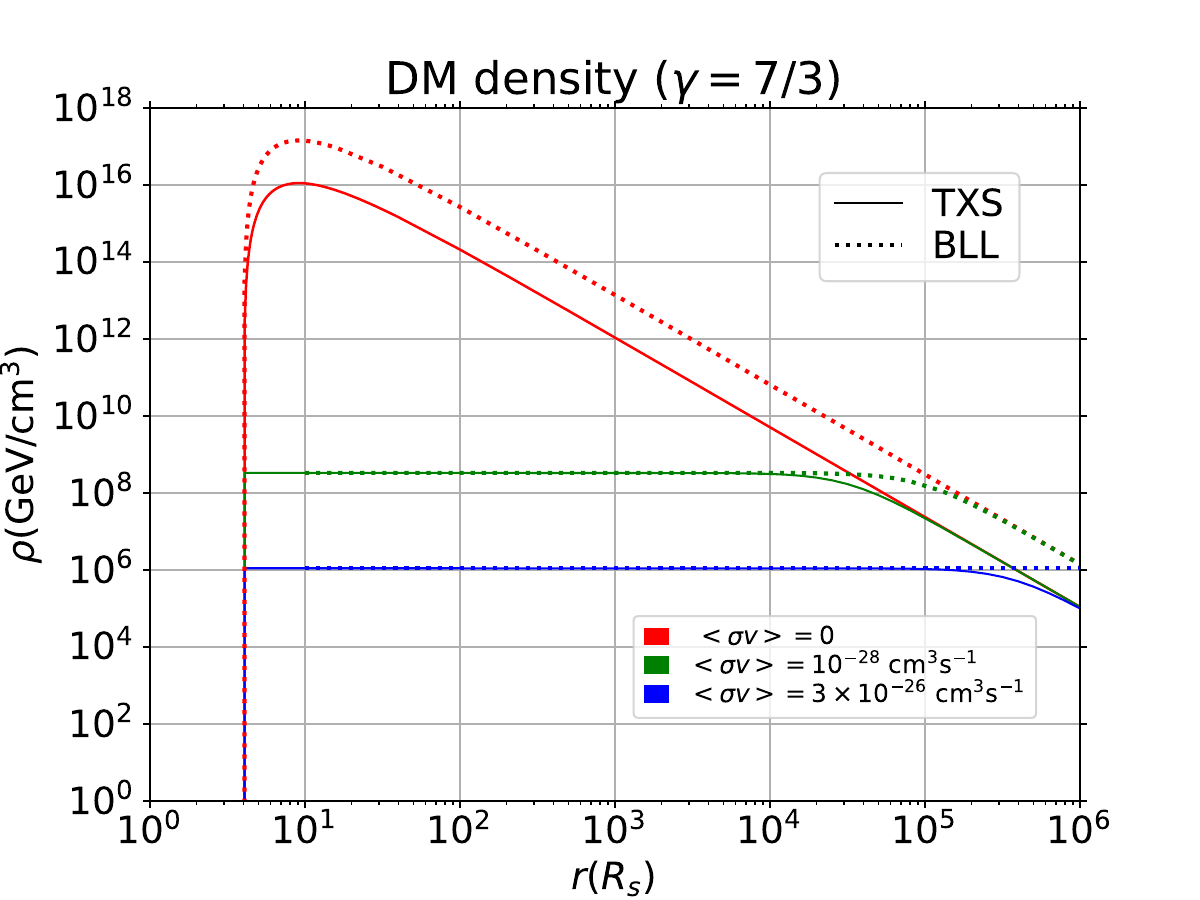}}
    \subfigure[\label{den2}]{
    \includegraphics[scale=0.4]{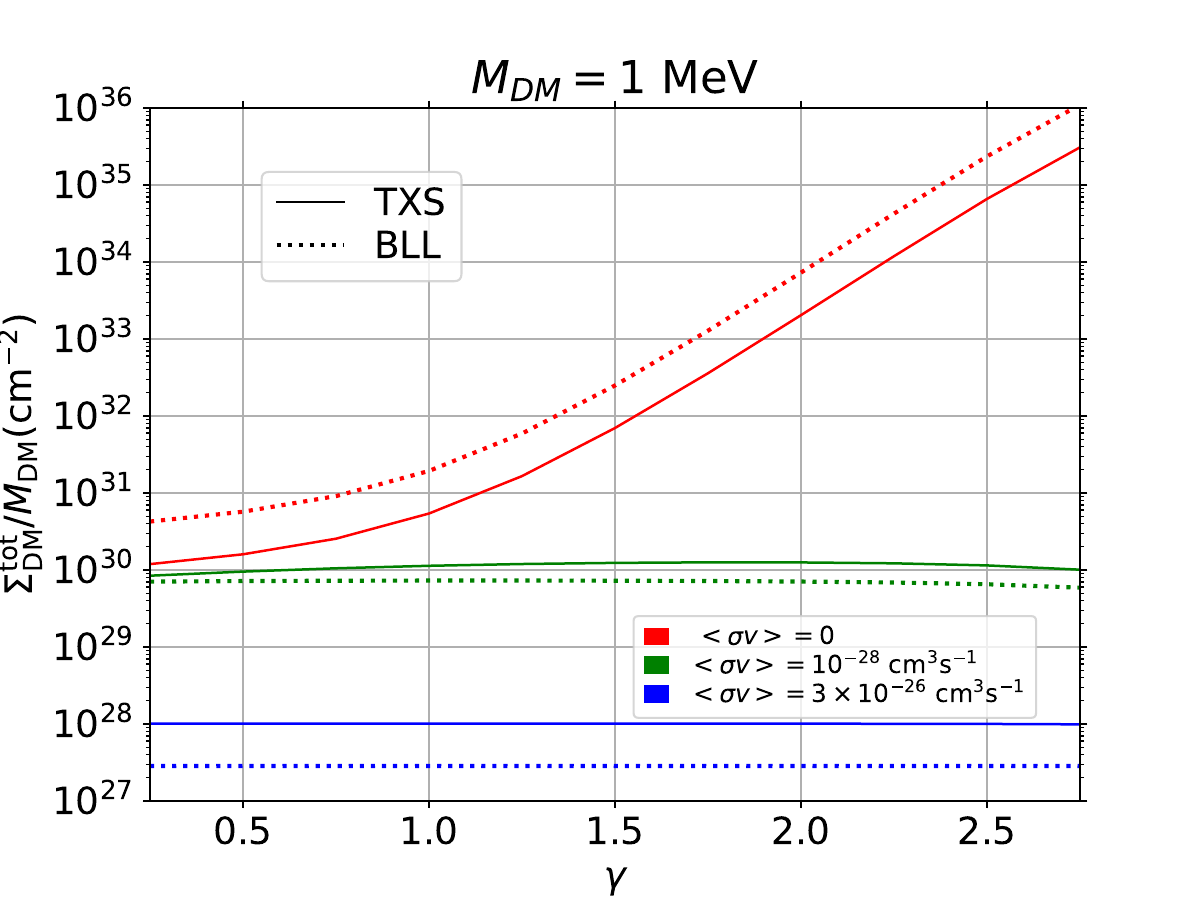}}
    \caption{ DM density profile: (a) Variation in $\rho_{\rm DM}$ with distance from BH center (in terms of $R_s$) with $\gamma=7/3$ for TXS 0506$+56$ and BL Lacertae
shown by solid and dotted lines. (b) Variation of $\Sigma_{\rm DM}^{\rm tot}/m_\chi$ with $\gamma$ for a fixed DM mass 1 MeV for the earlier mentioned two BHs.
In both the plots the red, green and blue line correspond to the cases with $\langle \sigma v\rangle = 0, 10^{-28}$ cm$^3$s$^{-1},~ 3\times 10^{-26}$ cm$^3$s$^{-1}$.}
    \label{fig:density}
\end{figure}
In Fig.\ref{den1} we show the variation of DM density with distance from BH center (in terms of $R_s$) with $\gamma=7/3$ as used in literature \cite{Wang:2021jic,Granelli:2022ysi}.
The {solid and dotted lines represent} DM profile due to  TXS 0506$+56$ and BL Lacertae respectively.
The red, green and blue line correspond to the cases with $\langle \sigma v\rangle = 0, 10^{-28}$ cm$^3$s$^{-1},~ 3\times 10^{-26}$ cm$^3$s$^{-1}$.
Note that the annihilation dilutes DM density and flattens the spike towards a core profile as expected from the discussion above.
Following the same argument with an increase in $\langle \sigma v\rangle$ the density decreases. 

In Fig.\ref{den2} we present the variation of $\Sigma_{\rm DM}^{\rm tot}/m_\chi$ with $\gamma$ for a fixed DM mass 1 MeV {for the two BHs mentioned above}. The line type and colors follow the same convention as in Fig.\ref{den1}. 
Note that, the spike radius is controlled by the BH mass ($R_s(\propto M_{BH})$) and hence it is larger for TXS 0506$+56$ than BL Lacertae due to the heavier mass of the former one. 
For the same reason normalization in density $\rho_1$ ( $\propto M_{BH}^{\gamma-2}$) and $\Sigma_{\rm DM}^{\rm tot}(\propto M_{BH}^{-1})$ is smaller for TXS 0506$+56$ than BL Lacertae. 
This feature is reflected in Fig.\ref{den2} where BL Lacertae has larger LOS density for $\langle\sigma v\rangle=0$.
Note that for $\langle \sigma v\rangle \neq 0$ $\Sigma_{\rm DM}^{\rm tot}$ doesn't vary significantly with $\gamma$.
However, for $\langle \sigma v\rangle=0$, $\Sigma_{\rm DM}^{\rm tot}$ varies by order's of magnitude with $\gamma$.
With an increase in $\gamma$ the DM spike becomes steeper leading to higher density as reflected in the aforementioned plot.

%Having a detailed discussion about the blazar jet and DM spike density we are now set to evaluate the boosted DM flux.

%&%%%%%%%%%%%%%%%%%%%%%%%%%%%%%%%%%%
\section{Existing constraints on ALP mediator}
\label{apx:constraint}
%==============
For a mediator coupled with electrons stringent bounds come from several ground based experiments as well as from astrophysics. We enumerate them below. Though we specifically discuss here about ALP mediator, the process of obtaining the bounds and the qualitative features remain same for vector mediator as well.
\begin{itemize}
    \item {\bf LEP:} Large electron positron collider (LEP) at CERN searches for missing energy through the process $e^++e^-\to \gamma+X,~X$ being the missing energy \cite{Alekseev:2001va}, and thus excludes $g_{ae}\gtrsim 2 \times 10^{-2}$ for $m_a> 2 m_\chi$ \cite{Angel:2023exb}. 
    \item {\bf BaBar:} BaBar also looks for missing energy through the same process as LEP does and 
    places stringent constraint on $g_{ae}\gtrsim3 \times 10^{-4} $ for $m_a$ upto 10 GeV \cite{BaBar:2014zli,Bauer:2017ris,Liu:2023bby,Armando:2023zwz,Angel:2023exb}. 
    \item {\bf NA64:} Though NA64 is one type of Beam Dump experiment, we discuss separately as it sets the most stringent limit on the parameter space relevant to us. In contrast to previous experiments NA64 looks for both visible and invisible decay of the mediator \cite{NA64:2021aiq,Andreev:2021fzd}. Here high energy electron beam is dumped at target like 
    $e^-+Z\to e^-+Z+a^*$, and then the mediator can decay either visibly or invisibly.
    In the visible channel ALP mediator can decay to SM particles (di-photon via loop for $m_a<2 m_e$) and no excess in observed data leads to constraint on $g_{ae}\gtrsim 10^{-5}$\cite{NA64:2020qwq}.
    On the other hand the invisible decay sets limit for $1$ MeV $<m_a<15$ MeV for $m_a> 2m_\chi$ \cite{NA64:2021aiq}.
    \item {\bf Beam Dump:} In this paper by ``Beam Dump" we refer to other fixed target experiments looking for leptonic decay of the mediator through displaced vertex search like E137, E141, Orsay \cite{Bjorken:1988as,Bechis:1979kp,Bechis:1979kp}. These experiments constrain roughly $g_{ae}\gtrsim 10^{-7}$ for  $1$ MeV $<m_a<100$ MeV \cite{Blumlein:1991xh,Andreas:2010ms,Liu:2017htz,Waites:2022tov} .
    \item {\bf Astrophysical constraints:} The most studied astrophysical constraint on light mediator is from supernova cooling. Light mediators with mass $\lesssim \mathcal{O}$(MeV) can be produced in the core and with feeble interaction it may escape the core leading to cooling of supernova. The observation of SN1987A provides the information about the luminosity and constrains such energy loss placing constraint on $g_{ae}$ \cite{Carenza:2021pcm} . However SN1987A bounds are disputed in literature due to the uncertainties associated with the progenitor \cite{Bar:2019ifz}. Nevertheless these bounds are applicable to very small couplings ($g_{ae}\sim 10^{-7}$) compared to our region of interest. Recent study has also pointed out strong constraint on leptophilic scalar mediator mass upto $\sim 10$ MeV \cite{Hardy:2024gwy}. Translating these bounds for ALP mediator is beyond the scope of this work. However, even after considering this bound we still have sufficient parameter space for $m_a\gtrsim10$ MeV. Similarly other astrophysical bodies like red giant, white dwarf places constraint for $m_a\lesssim \mathcal{O}(1)$ keV \cite{Giannotti:2015kwo}.
\end{itemize}
It is also worth discussing the dependence of the constraints on DM mass. For our minimal model the invisible decay constraints coming from BaBar, NA64 invisible are relevant as long as mediator mass $>2 m_\chi$ so that it can decay dominantly to dark sector as mentioned earlier. 
On the other hand the visible decay constraints (Beam Dump, NA64 visible) are derived assuming that  the mediator decays to SM particles with 100$\%$ branching ratio (BR).

\bibliography{ref}

\providecommand{\href}[2]{#2}\begingroup\raggedright\begin{thebibliography}{10}

\bibitem{Zwicky:1933gu}
F.~Zwicky, \emph{{Die Rotverschiebung von extragalaktischen Nebeln}},
  \href{https://doi.org/10.1007/s10714-008-0707-4}{\emph{Helv. Phys. Acta}
  {\bfseries 6} (1933) 110}.

\bibitem{Rubin:1970zza}
V.C.~Rubin and W.K.~Ford, Jr., \emph{{Rotation of the Andromeda Nebula from a
  Spectroscopic Survey of Emission Regions}},
  \href{https://doi.org/10.1086/150317}{\emph{Astrophys. J.} {\bfseries 159}
  (1970) 379}.

\bibitem{PAMELA:2011bbe}
{\scshape PAMELA} collaboration, \emph{{The cosmic-ray electron flux measured
  by the PAMELA experiment between 1 and 625 GeV}},
  \href{https://doi.org/10.1103/PhysRevLett.106.201101}{\emph{Phys. Rev. Lett.}
  {\bfseries 106} (2011) 201101}
  [\href{https://arxiv.org/abs/1103.2880}{{\ttfamily 1103.2880}}].

\bibitem{Planck:2018vyg}
{\scshape Planck} collaboration, \emph{{Planck 2018 results. VI. Cosmological
  parameters}},
  \href{https://doi.org/10.1051/0004-6361/201833910}{\emph{Astron. Astrophys.}
  {\bfseries 641} (2020) A6}
  [\href{https://arxiv.org/abs/1807.06209}{{\ttfamily 1807.06209}}].

\bibitem{Scherrer:1985zt}
R.J.~Scherrer and M.S.~Turner, \emph{{On the Relic, Cosmic Abundance of Stable
  Weakly Interacting Massive Particles}},
  \href{https://doi.org/10.1103/PhysRevD.33.1585}{\emph{Phys. Rev. D}
  {\bfseries 33} (1986) 1585}.

\bibitem{Jungman:1995df}
G.~Jungman, M.~Kamionkowski and K.~Griest, \emph{{Supersymmetric dark matter}},
  \href{https://doi.org/10.1016/0370-1573(95)00058-5}{\emph{Phys. Rept.}
  {\bfseries 267} (1996) 195}
  [\href{https://arxiv.org/abs/hep-ph/9506380}{{\ttfamily hep-ph/9506380}}].

\bibitem{Cirelli:2024ssz}
M.~Cirelli, A.~Strumia and J.~Zupan, \emph{{Dark Matter}},
  \href{https://arxiv.org/abs/2406.01705}{{\ttfamily 2406.01705}}.

\bibitem{Arcadi:2017kky}
G.~Arcadi, M.~Dutra, P.~Ghosh, M.~Lindner, Y.~Mambrini, M.~Pierre et~al.,
  \emph{{The waning of the WIMP? A review of models, searches, and
  constraints}},
  \href{https://doi.org/10.1140/epjc/s10052-018-5662-y}{\emph{Eur. Phys. J. C}
  {\bfseries 78} (2018) 203}
  [\href{https://arxiv.org/abs/1703.07364}{{\ttfamily 1703.07364}}].

\bibitem{Feng:2010gw}
J.L.~Feng, \emph{{Dark Matter Candidates from Particle Physics and Methods of
  Detection}},
  \href{https://doi.org/10.1146/annurev-astro-082708-101659}{\emph{Ann. Rev.
  Astron. Astrophys.} {\bfseries 48} (2010) 495}
  [\href{https://arxiv.org/abs/1003.0904}{{\ttfamily 1003.0904}}].

\bibitem{Lin:2019uvt}
T.~Lin, \emph{{Dark matter models and direct detection}},
  \href{https://doi.org/10.22323/1.333.0009}{\emph{PoS} {\bfseries 333} (2019)
  009} [\href{https://arxiv.org/abs/1904.07915}{{\ttfamily 1904.07915}}].

\bibitem{XENON:2022ltv}
{\scshape XENON} collaboration, \emph{{Search for New Physics in Electronic
  Recoil Data from XENONnT}},
  \href{https://doi.org/10.1103/PhysRevLett.129.161805}{\emph{Phys. Rev. Lett.}
  {\bfseries 129} (2022) 161805}
  [\href{https://arxiv.org/abs/2207.11330}{{\ttfamily 2207.11330}}].

\bibitem{XENON:2023cxc}
{\scshape XENON} collaboration, \emph{{First Dark Matter Search with Nuclear
  Recoils from the XENONnT Experiment}},
  \href{https://doi.org/10.1103/PhysRevLett.131.041003}{\emph{Phys. Rev. Lett.}
  {\bfseries 131} (2023) 041003}
  [\href{https://arxiv.org/abs/2303.14729}{{\ttfamily 2303.14729}}].

\bibitem{LZ:2023poo}
{\scshape LZ} collaboration, \emph{{Search for new physics in low-energy
  electron recoils from the first LZ exposure}},
  \href{https://doi.org/10.1103/PhysRevD.108.072006}{\emph{Phys. Rev. D}
  {\bfseries 108} (2023) 072006}
  [\href{https://arxiv.org/abs/2307.15753}{{\ttfamily 2307.15753}}].

\bibitem{LZCollaboration:2024lux}
{\scshape LZ Collaboration} collaboration, \emph{{Dark Matter Search Results
  from 4.2 Tonne-Years of Exposure of the LUX-ZEPLIN (LZ) Experiment}},
  \href{https://arxiv.org/abs/2410.17036}{{\ttfamily 2410.17036}}.

\bibitem{PandaX-II:2020oim}
{\scshape PandaX-II} collaboration, \emph{{Results of dark matter search using
  the full PandaX-II exposure}},
  \href{https://doi.org/10.1088/1674-1137/abb658}{\emph{Chin. Phys. C}
  {\bfseries 44} (2020) 125001}
  [\href{https://arxiv.org/abs/2007.15469}{{\ttfamily 2007.15469}}].

\bibitem{LUX:2016ggv}
{\scshape LUX} collaboration, \emph{{Results from a search for dark matter in
  the complete LUX exposure}},
  \href{https://doi.org/10.1103/PhysRevLett.118.021303}{\emph{Phys. Rev. Lett.}
  {\bfseries 118} (2017) 021303}
  [\href{https://arxiv.org/abs/1608.07648}{{\ttfamily 1608.07648}}].

\bibitem{DEAP:2019yzn}
{\scshape DEAP} collaboration, \emph{{Search for dark matter with a 231-day
  exposure of liquid argon using DEAP-3600 at SNOLAB}},
  \href{https://doi.org/10.1103/PhysRevD.100.022004}{\emph{Phys. Rev. D}
  {\bfseries 100} (2019) 022004}
  [\href{https://arxiv.org/abs/1902.04048}{{\ttfamily 1902.04048}}].

\bibitem{DarkSide:2018kuk}
{\scshape DarkSide} collaboration, \emph{{DarkSide-50 532-day Dark Matter
  Search with Low-Radioactivity Argon}},
  \href{https://doi.org/10.1103/PhysRevD.98.102006}{\emph{Phys. Rev. D}
  {\bfseries 98} (2018) 102006}
  [\href{https://arxiv.org/abs/1802.07198}{{\ttfamily 1802.07198}}].

\bibitem{Yin:2018yjn}
W.~Yin, \emph{{Highly-boosted dark matter and cutoff for cosmic-ray neutrinos
  through neutrino portal}},
  \href{https://doi.org/10.1051/epjconf/201920804003}{\emph{EPJ Web Conf.}
  {\bfseries 208} (2019) 04003}
  [\href{https://arxiv.org/abs/1809.08610}{{\ttfamily 1809.08610}}].

\bibitem{Bringmann:2018cvk}
T.~Bringmann and M.~Pospelov, \emph{{Novel direct detection constraints on
  light dark matter}},
  \href{https://doi.org/10.1103/PhysRevLett.122.171801}{\emph{Phys. Rev. Lett.}
  {\bfseries 122} (2019) 171801}
  [\href{https://arxiv.org/abs/1810.10543}{{\ttfamily 1810.10543}}].

\bibitem{Das:2021lcr}
A.~Das and M.~Sen, \emph{{Boosted dark matter from diffuse supernova
  neutrinos}}, \href{https://doi.org/10.1103/PhysRevD.104.075029}{\emph{Phys.
  Rev. D} {\bfseries 104} (2021) 075029}
  [\href{https://arxiv.org/abs/2104.00027}{{\ttfamily 2104.00027}}].

\bibitem{Ghosh:2021vkt}
D.~Ghosh, A.~Guha and D.~Sachdeva, \emph{{Exclusion limits on dark
  matter-neutrino scattering cross section}},
  \href{https://doi.org/10.1103/PhysRevD.105.103029}{\emph{Phys. Rev. D}
  {\bfseries 105} (2022) 103029}
  [\href{https://arxiv.org/abs/2110.00025}{{\ttfamily 2110.00025}}].

\bibitem{Ema:2018bih}
Y.~Ema, F.~Sala and R.~Sato, \emph{{Light Dark Matter at Neutrino
  Experiments}},
  \href{https://doi.org/10.1103/PhysRevLett.122.181802}{\emph{Phys. Rev. Lett.}
  {\bfseries 122} (2019) 181802}
  [\href{https://arxiv.org/abs/1811.00520}{{\ttfamily 1811.00520}}].

\bibitem{Maity:2022exk}
T.N.~Maity and R.~Laha, \emph{{Cosmic-ray boosted dark matter in Xe-based
  direct detection experiments}},
  \href{https://doi.org/10.1140/epjc/s10052-024-12464-8}{\emph{Eur. Phys. J. C}
  {\bfseries 84} (2024) 117}
  [\href{https://arxiv.org/abs/2210.01815}{{\ttfamily 2210.01815}}].

\bibitem{DeRomeri:2023ytt}
V.~De~Romeri, A.~Majumdar, D.K.~Papoulias and R.~Srivastava, \emph{{XENONnT and
  LUX-ZEPLIN constraints on DSNB-boosted dark matter}},
  \href{https://doi.org/10.1088/1475-7516/2024/03/028}{\emph{JCAP} {\bfseries
  03} (2024) 028} [\href{https://arxiv.org/abs/2309.04117}{{\ttfamily
  2309.04117}}].

\bibitem{Bardhan:2022bdg}
D.~Bardhan, S.~Bhowmick, D.~Ghosh, A.~Guha and D.~Sachdeva, \emph{{Bounds on
  boosted dark matter from direct detection: The role of energy-dependent cross
  sections}}, \href{https://doi.org/10.1103/PhysRevD.107.015010}{\emph{Phys.
  Rev. D} {\bfseries 107} (2023) 015010}
  [\href{https://arxiv.org/abs/2208.09405}{{\ttfamily 2208.09405}}].

\bibitem{Dent:2019krz}
J.B.~Dent, B.~Dutta, J.L.~Newstead and I.M.~Shoemaker, \emph{{Bounds on Cosmic
  Ray-Boosted Dark Matter in Simplified Models and its Corresponding
  Neutrino-Floor}},
  \href{https://doi.org/10.1103/PhysRevD.101.116007}{\emph{Phys. Rev. D}
  {\bfseries 101} (2020) 116007}
  [\href{https://arxiv.org/abs/1907.03782}{{\ttfamily 1907.03782}}].

\bibitem{Ema:2020ulo}
Y.~Ema, F.~Sala and R.~Sato, \emph{{Neutrino experiments probe hadrophilic
  light dark matter}},
  \href{https://doi.org/10.21468/SciPostPhys.10.3.072}{\emph{SciPost Phys.}
  {\bfseries 10} (2021) 072}
  [\href{https://arxiv.org/abs/2011.01939}{{\ttfamily 2011.01939}}].

\bibitem{Bell:2023sdq}
N.F.~Bell, J.L.~Newstead and I.~Shaukat-Ali, \emph{{Cosmic-ray boosted dark
  matter confronted by constraints on new light mediators}},
  \href{https://doi.org/10.1103/PhysRevD.109.063034}{\emph{Phys. Rev. D}
  {\bfseries 109} (2024) 063034}
  [\href{https://arxiv.org/abs/2309.11003}{{\ttfamily 2309.11003}}].

\bibitem{Guha:2024mjr}
A.~Guha and J.-C.~Park, \emph{{Constraints on cosmic-ray boosted dark matter
  with realistic cross section}},
  \href{https://doi.org/10.1088/1475-7516/2024/07/074}{\emph{JCAP} {\bfseries
  07} (2024) 074} [\href{https://arxiv.org/abs/2401.07750}{{\ttfamily
  2401.07750}}].

\bibitem{Wang:2021jic}
J.-W.~Wang, A.~Granelli and P.~Ullio, \emph{{Direct Detection Constraints on
  Blazar-Boosted Dark Matter}},
  \href{https://doi.org/10.1103/PhysRevLett.128.221104}{\emph{Phys. Rev. Lett.}
  {\bfseries 128} (2022) 221104}
  [\href{https://arxiv.org/abs/2111.13644}{{\ttfamily 2111.13644}}].

\bibitem{Granelli:2022ysi}
A.~Granelli, P.~Ullio and J.-W.~Wang, \emph{{Blazar-boosted dark matter at
  Super-Kamiokande}},
  \href{https://doi.org/10.1088/1475-7516/2022/07/013}{\emph{JCAP} {\bfseries
  07} (2022) 013} [\href{https://arxiv.org/abs/2202.07598}{{\ttfamily
  2202.07598}}].

\bibitem{Urry:1995mg}
C.M.~Urry and P.~Padovani, \emph{{Unified schemes for radio-loud active
  galactic nuclei}}, \href{https://doi.org/10.1086/133630}{\emph{Publ. Astron.
  Soc. Pac.} {\bfseries 107} (1995) 803}
  [\href{https://arxiv.org/abs/astro-ph/9506063}{{\ttfamily
  astro-ph/9506063}}].

\bibitem{Abdo_2010}
A.A.~Abdo, M.~Ackermann, I.~Agudo, M.~Ajello, H.D.~Aller, M.F.~Aller et~al.,
  \emph{The spectral energy distribution offermibright blazars},
  \href{https://doi.org/10.1088/0004-637x/716/1/30}{\emph{The Astrophysical
  Journal} {\bfseries 716} (2010) 30–70}.

\bibitem{Gondolo:1999ef}
P.~Gondolo and J.~Silk, \emph{{Dark matter annihilation at the galactic
  center}}, \href{https://doi.org/10.1103/PhysRevLett.83.1719}{\emph{Phys. Rev.
  Lett.} {\bfseries 83} (1999) 1719}
  [\href{https://arxiv.org/abs/astro-ph/9906391}{{\ttfamily
  astro-ph/9906391}}].

\bibitem{Bhowmick:2022zkj}
S.~Bhowmick, D.~Ghosh and D.~Sachdeva, \emph{{Blazar boosted dark matter
  \textemdash{} direct detection constraints on
  \ensuremath{\sigma}e\ensuremath{\chi}: role of energy dependent cross
  sections}}, \href{https://doi.org/10.1088/1475-7516/2023/07/039}{\emph{JCAP}
  {\bfseries 07} (2023) 039}
  [\href{https://arxiv.org/abs/2301.00209}{{\ttfamily 2301.00209}}].

\bibitem{DeMarchi:2024riu}
A.G.~De~Marchi, A.~Granelli, J.~Nava and F.~Sala, \emph{{Did IceCube discover
  Dark Matter around Blazars?}},
  \href{https://arxiv.org/abs/2412.07861}{{\ttfamily 2412.07861}}.

\bibitem{Herrera:2023nww}
G.~Herrera and K.~Murase, \emph{{Probing light dark matter through cosmic-ray
  cooling in active galactic nuclei}},
  \href{https://doi.org/10.1103/PhysRevD.110.L011701}{\emph{Phys. Rev. D}
  {\bfseries 110} (2024) L011701}
  [\href{https://arxiv.org/abs/2307.09460}{{\ttfamily 2307.09460}}].

\bibitem{Super-Kamiokande:2017dch}
{\scshape Super-Kamiokande} collaboration, \emph{{Search for Boosted Dark
  Matter Interacting With Electrons in Super-Kamiokande}},
  \href{https://doi.org/10.1103/PhysRevLett.120.221301}{\emph{Phys. Rev. Lett.}
  {\bfseries 120} (2018) 221301}
  [\href{https://arxiv.org/abs/1711.05278}{{\ttfamily 1711.05278}}].

\bibitem{Peccei:1977hh}
R.D.~Peccei and H.R.~Quinn, \emph{{CP Conservation in the Presence of
  Instantons}}, \href{https://doi.org/10.1103/PhysRevLett.38.1440}{\emph{Phys.
  Rev. Lett.} {\bfseries 38} (1977) 1440}.

\bibitem{Bharucha:2022lty}
A.~Bharucha, F.~Br\"ummer, N.~Desai and S.~Mutzel, \emph{{Axion-like particles
  as mediators for dark matter: beyond freeze-out}},
  \href{https://doi.org/10.1007/JHEP02(2023)141}{\emph{JHEP} {\bfseries 02}
  (2023) 141} [\href{https://arxiv.org/abs/2209.03932}{{\ttfamily
  2209.03932}}].

\bibitem{Ghosh:2023tyz}
D.K.~Ghosh, A.~Ghoshal and S.~Jeesun, \emph{{Axion-like particle (ALP) portal
  freeze-in dark matter confronting ALP search experiments}},
  \href{https://doi.org/10.1007/JHEP01(2024)026}{\emph{JHEP} {\bfseries 01}
  (2024) 026} [\href{https://arxiv.org/abs/2305.09188}{{\ttfamily
  2305.09188}}].

\bibitem{Bauer:2017ris}
M.~Bauer, M.~Neubert and A.~Thamm, \emph{{Collider Probes of Axion-Like
  Particles}}, \href{https://doi.org/10.1007/JHEP12(2017)044}{\emph{JHEP}
  {\bfseries 12} (2017) 044}
  [\href{https://arxiv.org/abs/1708.00443}{{\ttfamily 1708.00443}}].

\bibitem{NA64:2021aiq}
{\scshape NA64} collaboration, \emph{{Search for pseudoscalar bosons decaying
  into $e^+e^-$ pairs in the NA64 experiment at the CERN SPS}},
  \href{https://doi.org/10.1103/PhysRevD.104.L111102}{\emph{Phys. Rev. D}
  {\bfseries 104} (2021) L111102}
  [\href{https://arxiv.org/abs/2104.13342}{{\ttfamily 2104.13342}}].

\bibitem{Cerruti:2018tmc}
M.~Cerruti, A.~Zech, C.~Boisson, G.~Emery, S.~Inoue and J.P.~Lenain,
  \emph{{Leptohadronic single-zone models for the electromagnetic and neutrino
  emission of TXS 0506+056}},
  \href{https://doi.org/10.1093/mnrasl/sly210}{\emph{Mon. Not. Roy. Astron.
  Soc.} {\bfseries 483} (2019) L12}
  [\href{https://arxiv.org/abs/1807.04335}{{\ttfamily 1807.04335}}].

\bibitem{Boettcher:2013wxa}
M.~Boettcher, A.~Reimer, K.~Sweeney and A.~Prakash, \emph{{Leptonic and
  Hadronic Modeling of Fermi-Detected Blazars}},
  \href{https://doi.org/10.1088/0004-637X/768/1/54}{\emph{Astrophys. J.}
  {\bfseries 768} (2013) 54} [\href{https://arxiv.org/abs/1304.0605}{{\ttfamily
  1304.0605}}].

\bibitem{IceCube:2018dnn}
{\scshape IceCube, Fermi-LAT, MAGIC, AGILE, ASAS-SN, HAWC, H.E.S.S., INTEGRAL,
  Kanata, Kiso, Kapteyn, Liverpool Telescope, Subaru, Swift NuSTAR, VERITAS,
  VLA/17B-403} collaboration, \emph{{Multimessenger observations of a flaring
  blazar coincident with high-energy neutrino IceCube-170922A}},
  \href{https://doi.org/10.1126/science.aat1378}{\emph{Science} {\bfseries 361}
  (2018) eaat1378} [\href{https://arxiv.org/abs/1807.08816}{{\ttfamily
  1807.08816}}].

\bibitem{Cerruti:2020lfj}
M.~Cerruti, \emph{{Leptonic and Hadronic Radiative Processes in
  Supermassive-Black-Hole Jets}},
  \href{https://doi.org/10.3390/galaxies8040072}{\emph{Galaxies} {\bfseries 8}
  (2020) 72} [\href{https://arxiv.org/abs/2012.13302}{{\ttfamily 2012.13302}}].

\bibitem{Ghisellini_2009}
G.~Ghisellini, F.~Tavecchio, L.~Foschini, G.~Ghirlanda, L.~Maraschi and
  A.~Celotti, \emph{General physical properties of bright fermi blazars:
  Properties of bright fermi blazars},
  \href{https://doi.org/10.1111/j.1365-2966.2009.15898.x}{\emph{Monthly Notices
  of the Royal Astronomical Society} {\bfseries 402} (2009) 497–518}.

\bibitem{Celotti:2007rb}
A.~Celotti and G.~Ghisellini, \emph{{The power of blazar jets}},
  \href{https://doi.org/10.1111/j.1365-2966.2007.12758.x}{\emph{Mon. Not. Roy.
  Astron. Soc.} {\bfseries 385} (2008) 283}
  [\href{https://arxiv.org/abs/0711.4112}{{\ttfamily 0711.4112}}].

\bibitem{Gorchtein:2010xa}
M.~Gorchtein, S.~Profumo and L.~Ubaldi, \emph{{Probing Dark Matter with AGN
  Jets}}, \href{https://doi.org/10.1103/PhysRevD.82.083514}{\emph{Phys. Rev. D}
  {\bfseries 82} (2010) 083514}
  [\href{https://arxiv.org/abs/1008.2230}{{\ttfamily 1008.2230}}].

\bibitem{Ullio:2001fb}
P.~Ullio, H.~Zhao and M.~Kamionkowski, \emph{{A Dark matter spike at the
  galactic center?}},
  \href{https://doi.org/10.1103/PhysRevD.64.043504}{\emph{Phys. Rev. D}
  {\bfseries 64} (2001) 043504}
  [\href{https://arxiv.org/abs/astro-ph/0101481}{{\ttfamily
  astro-ph/0101481}}].

\bibitem{Gnedin:2003rj}
O.Y.~Gnedin and J.R.~Primack, \emph{{Dark Matter Profile in the Galactic
  Center}}, \href{https://doi.org/10.1103/PhysRevLett.93.061302}{\emph{Phys.
  Rev. Lett.} {\bfseries 93} (2004) 061302}
  [\href{https://arxiv.org/abs/astro-ph/0308385}{{\ttfamily
  astro-ph/0308385}}].

\bibitem{Liu:2023bby}
J.~Liu, Y.~Luo and M.~Song, \emph{{Investigation of the concurrent effects of
  ALP-photon and ALP-electron couplings in Collider and Beam Dump Searches}},
  \href{https://doi.org/10.1007/JHEP09(2023)104}{\emph{JHEP} {\bfseries 09}
  (2023) 104} [\href{https://arxiv.org/abs/2304.05435}{{\ttfamily
  2304.05435}}].

\bibitem{Armando:2023zwz}
G.~Armando, P.~Panci, J.~Weiss and R.~Ziegler, \emph{{Leptonic ALP portal to
  the dark sector}},
  \href{https://doi.org/10.1103/PhysRevD.109.055029}{\emph{Phys. Rev. D}
  {\bfseries 109} (2024) 055029}
  [\href{https://arxiv.org/abs/2310.05827}{{\ttfamily 2310.05827}}].

\bibitem{Angel:2023exb}
L.~Angel et~al., \emph{{Toward a search for axionlike particles at the LNLS}},
  \href{https://doi.org/10.1103/PhysRevD.108.055030}{\emph{Phys. Rev. D}
  {\bfseries 108} (2023) 055030}
  [\href{https://arxiv.org/abs/2305.13384}{{\ttfamily 2305.13384}}].

\bibitem{Bechis:1979kp}
D.J.~Bechis, T.W.~Dombeck, R.W.~Ellsworth, E.V.~Sager, P.H.~Steinberg,
  L.J.~Teig et~al., \emph{{Search for Axion Production in Low-energy Electron
  Bremsstrahlung}},
  \href{https://doi.org/10.1103/PhysRevLett.42.1511}{\emph{Phys. Rev. Lett.}
  {\bfseries 42} (1979) 1511}.

\bibitem{Blumlein:1991xh}
J.~Blumlein et~al., \emph{{Limits on the mass of light (pseudo)scalar particles
  from Bethe-Heitler e+ e- and mu+ mu- pair production in a proton - iron beam
  dump experiment}},
  \href{https://doi.org/10.1142/S0217751X9200171X}{\emph{Int. J. Mod. Phys. A}
  {\bfseries 7} (1992) 3835}.

\bibitem{Andreas:2010ms}
S.~Andreas, O.~Lebedev, S.~Ramos-Sanchez and A.~Ringwald, \emph{{Constraints on
  a very light CP-odd Higgs of the NMSSM and other axion-like particles}},
  \href{https://doi.org/10.1007/JHEP08(2010)003}{\emph{JHEP} {\bfseries 08}
  (2010) 003} [\href{https://arxiv.org/abs/1005.3978}{{\ttfamily 1005.3978}}].

\bibitem{Liu:2017htz}
Y.-S.~Liu and G.A.~Miller, \emph{{Validity of the Weizs\"acker-Williams
  approximation and the analysis of beam dump experiments: Production of an
  axion, a dark photon, or a new axial-vector boson}},
  \href{https://doi.org/10.1103/PhysRevD.96.016004}{\emph{Phys. Rev. D}
  {\bfseries 96} (2017) 016004}
  [\href{https://arxiv.org/abs/1705.01633}{{\ttfamily 1705.01633}}].

\bibitem{Waites:2022tov}
L.~Waites, A.~Thompson, A.~Bungau, J.M.~Conrad, B.~Dutta, W.-C.~Huang et~al.,
  \emph{{Axionlike particle production at beam dump experiments with distinct
  nuclear excitation lines}},
  \href{https://doi.org/10.1103/PhysRevD.107.095010}{\emph{Phys. Rev. D}
  {\bfseries 107} (2023) 095010}
  [\href{https://arxiv.org/abs/2207.13659}{{\ttfamily 2207.13659}}].

\bibitem{Andreev:2021fzd}
Y.M.~Andreev et~al., \emph{{Improved exclusion limit for light dark matter from
  e+e- annihilation in NA64}},
  \href{https://doi.org/10.1103/PhysRevD.104.L091701}{\emph{Phys. Rev. D}
  {\bfseries 104} (2021) L091701}
  [\href{https://arxiv.org/abs/2108.04195}{{\ttfamily 2108.04195}}].

\bibitem{Caputo:2024oqc}
A.~Caputo and G.~Raffelt, \emph{{Astrophysical Axion Bounds: The 2024
  Edition}}, \href{https://doi.org/10.22323/1.454.0041}{\emph{PoS} {\bfseries
  COSMICWISPers} (2024) 041}
  [\href{https://arxiv.org/abs/2401.13728}{{\ttfamily 2401.13728}}].

\bibitem{Tremaine:1979we}
S.~Tremaine and J.E.~Gunn, \emph{{Dynamical Role of Light Neutral Leptons in
  Cosmology}}, \href{https://doi.org/10.1103/PhysRevLett.42.407}{\emph{Phys.
  Rev. Lett.} {\bfseries 42} (1979) 407}.

\bibitem{Petraki:2013wwa}
K.~Petraki and R.R.~Volkas, \emph{{Review of asymmetric dark matter}},
  \href{https://doi.org/10.1142/S0217751X13300287}{\emph{Int. J. Mod. Phys. A}
  {\bfseries 28} (2013) 1330028}
  [\href{https://arxiv.org/abs/1305.4939}{{\ttfamily 1305.4939}}].

\bibitem{Okada:2020cue}
N.~Okada, S.~Okada and Q.~Shafi, \emph{{Light $Z'$ and dark matter from
  $U(1)_X$ gauge symmetry}},
  \href{https://doi.org/10.1016/j.physletb.2020.135845}{\emph{Phys. Lett. B}
  {\bfseries 810} (2020) 135845}
  [\href{https://arxiv.org/abs/2003.02667}{{\ttfamily 2003.02667}}].

\bibitem{Coloma:2022umy}
P.~Coloma, P.~Coloma, M.C.~Gonzalez-Garcia, M.C.~Gonzalez-Garcia, M.~Maltoni,
  M.~Maltoni et~al., \emph{{Constraining new physics with Borexino Phase-II
  spectral data}}, \href{https://doi.org/10.1007/JHEP07(2022)138}{\emph{JHEP}
  {\bfseries 07} (2022) 138}
  [\href{https://arxiv.org/abs/2204.03011}{{\ttfamily 2204.03011}}].

\bibitem{BaBar:2014zli}
{\scshape BaBar} collaboration, \emph{{Search for a Dark Photon in $e^+e^-$
  Collisions at BaBar}},
  \href{https://doi.org/10.1103/PhysRevLett.113.201801}{\emph{Phys. Rev. Lett.}
  {\bfseries 113} (2014) 201801}
  [\href{https://arxiv.org/abs/1406.2980}{{\ttfamily 1406.2980}}].

\bibitem{Bauer:2018onh}
M.~Bauer, P.~Foldenauer and J.~Jaeckel, \emph{{Hunting All the Hidden
  Photons}}, \href{https://doi.org/10.1007/JHEP07(2018)094}{\emph{JHEP}
  {\bfseries 07} (2018) 094}
  [\href{https://arxiv.org/abs/1803.05466}{{\ttfamily 1803.05466}}].

\bibitem{Banerjee:2019pds}
D.~Banerjee et~al., \emph{{Dark matter search in missing energy events with
  NA64}}, \href{https://doi.org/10.1103/PhysRevLett.123.121801}{\emph{Phys.
  Rev. Lett.} {\bfseries 123} (2019) 121801}
  [\href{https://arxiv.org/abs/1906.00176}{{\ttfamily 1906.00176}}].

\bibitem{Hardy:2016kme}
E.~Hardy and R.~Lasenby, \emph{{Stellar cooling bounds on new light particles:
  plasma mixing effects}},
  \href{https://doi.org/10.1007/JHEP02(2017)033}{\emph{JHEP} {\bfseries 02}
  (2017) 033} [\href{https://arxiv.org/abs/1611.05852}{{\ttfamily
  1611.05852}}].

\bibitem{Ghosh:2024cxi}
D.K.~Ghosh, P.~Ghosh, S.~Jeesun and R.~Srivastava, \emph{{The $N_{\rm eff}$ at
  CMB challenges $U(1)_X$ light gauge boson scenarios}},
  \href{https://arxiv.org/abs/2404.10077}{{\ttfamily 2404.10077}}.

\bibitem{Majumdar:2021vdw}
A.~Majumdar, D.K.~Papoulias and R.~Srivastava, \emph{{Dark matter detectors as
  a novel probe for light new physics}},
  \href{https://doi.org/10.1103/PhysRevD.106.013001}{\emph{Phys. Rev. D}
  {\bfseries 106} (2022) 013001}
  [\href{https://arxiv.org/abs/2112.03309}{{\ttfamily 2112.03309}}].

\bibitem{DeRomeri:2024dbv}
V.~De~Romeri, D.K.~Papoulias and C.A.~Ternes, \emph{{Light vector mediators at
  direct detection experiments}},
  \href{https://arxiv.org/abs/2402.05506}{{\ttfamily 2402.05506}}.

\bibitem{Khan:2019jvr}
A.N.~Khan, W.~Rodejohann and X.-J.~Xu, \emph{{Borexino and general neutrino
  interactions}},
  \href{https://doi.org/10.1103/PhysRevD.101.055047}{\emph{Phys. Rev. D}
  {\bfseries 101} (2020) 055047}
  [\href{https://arxiv.org/abs/1906.12102}{{\ttfamily 1906.12102}}].

\bibitem{AtzoriCorona:2022moj}
M.~Atzori~Corona, M.~Cadeddu, N.~Cargioli, F.~Dordei, C.~Giunti, Y.F.~Li
  et~al., \emph{{Probing light mediators and $(g-2)_{\mu}$ through detection of
  coherent elastic neutrino nucleus scattering at COHERENT}},
  \href{https://doi.org/10.1007/JHEP05(2022)109}{\emph{JHEP} {\bfseries 05}
  (2022) 109} [\href{https://arxiv.org/abs/2202.11002}{{\ttfamily
  2202.11002}}].

\bibitem{Ibe:2019gpv}
M.~Ibe, S.~Kobayashi, Y.~Nakayama and S.~Shirai, \emph{{Cosmological constraint
  on dark photon from N$_{eff}$}},
  \href{https://doi.org/10.1007/JHEP04(2020)009}{\emph{JHEP} {\bfseries 04}
  (2020) 009} [\href{https://arxiv.org/abs/1912.12152}{{\ttfamily
  1912.12152}}].

\bibitem{Escudero:2018mvt}
M.~Escudero, \emph{{Neutrino decoupling beyond the Standard Model: CMB
  constraints on the Dark Matter mass with a fast and precise $N_{\rm eff}$
  evaluation}},
  \href{https://doi.org/10.1088/1475-7516/2019/02/007}{\emph{JCAP} {\bfseries
  02} (2019) 007} [\href{https://arxiv.org/abs/1812.05605}{{\ttfamily
  1812.05605}}].

\bibitem{10.21468/SciPostPhys.10.3.072}
Y.~Ema, F.~Sala and R.~Sato, \emph{{Neutrino experiments probe hadrophilic
  light dark matter}},
  \href{https://doi.org/10.21468/SciPostPhys.10.3.072}{\emph{SciPost Phys.}
  {\bfseries 10} (2021) 072}.

\bibitem{Dolan:2014ska}
M.J.~Dolan, F.~Kahlhoefer, C.~McCabe and K.~Schmidt-Hoberg, \emph{{A taste of
  dark matter: Flavour constraints on pseudoscalar mediators}},
  \href{https://doi.org/10.1007/JHEP03(2015)171}{\emph{JHEP} {\bfseries 03}
  (2015) 171} [\href{https://arxiv.org/abs/1412.5174}{{\ttfamily 1412.5174}}].

\bibitem{Bjorken:1988as}
J.D.~Bjorken, S.~Ecklund, W.R.~Nelson, A.~Abashian, C.~Church, B.~Lu et~al.,
  \emph{{Search for Neutral Metastable Penetrating Particles Produced in the
  SLAC Beam Dump}}, \href{https://doi.org/10.1103/PhysRevD.38.3375}{\emph{Phys.
  Rev. D} {\bfseries 38} (1988) 3375}.

\bibitem{Okada:2018ktp}
S.~Okada, \emph{{$Z'$ Portal Dark Matter in the Minimal $B-L$ Model}},
  \href{https://doi.org/10.1155/2018/5340935}{\emph{Adv. High Energy Phys.}
  {\bfseries 2018} (2018) 5340935}
  [\href{https://arxiv.org/abs/1803.06793}{{\ttfamily 1803.06793}}].

\bibitem{Potter:2015tya}
W.J.~Potter and G.~Cotter, \emph{{New constraints on the structure and dynamics
  of black hole jets}}, \href{https://doi.org/10.1093/mnras/stv1657}{\emph{Mon.
  Not. Roy. Astron. Soc.} {\bfseries 453} (2015) 4070}
  [\href{https://arxiv.org/abs/1508.00567}{{\ttfamily 1508.00567}}].

\bibitem{refId0}
G.S.~Vila, G.E.~Romero and N.A.~Casco, \emph{An inhomogeneous lepto-hadronic
  model for the radiation of relativistic jets - application to xte j1118+480},
  \href{https://doi.org/10.1051/0004-6361/201118106}{\emph{A$\&$A} {\bfseries
  538} (2012) A97}.

\bibitem{Alekseev:2001va}
G.D.~Alekseev et~al., \emph{{The DELPHI experiment at LEP}}, {\emph{Part. Nucl.
  Lett.} {\bfseries 98} (2001) 5}.

\bibitem{NA64:2020qwq}
{\scshape NA64} collaboration, \emph{{Search for Axionlike and Scalar Particles
  with the NA64 Experiment}},
  \href{https://doi.org/10.1103/PhysRevLett.125.081801}{\emph{Phys. Rev. Lett.}
  {\bfseries 125} (2020) 081801}
  [\href{https://arxiv.org/abs/2005.02710}{{\ttfamily 2005.02710}}].

\bibitem{Carenza:2021pcm}
P.~Carenza and G.~Lucente, \emph{{Supernova bound on axionlike particles
  coupled with electrons}},
  \href{https://doi.org/10.1103/PhysRevD.104.103007}{\emph{Phys. Rev. D}
  {\bfseries 104} (2021) 103007}
  [\href{https://arxiv.org/abs/2107.12393}{{\ttfamily 2107.12393}}].

\bibitem{Bar:2019ifz}
N.~Bar, K.~Blum and G.~D'Amico, \emph{{Is there a supernova bound on axions?}},
  \href{https://doi.org/10.1103/PhysRevD.101.123025}{\emph{Phys. Rev. D}
  {\bfseries 101} (2020) 123025}
  [\href{https://arxiv.org/abs/1907.05020}{{\ttfamily 1907.05020}}].

\bibitem{Hardy:2024gwy}
E.~Hardy, A.~Sokolov and H.~Stubbs, \emph{{Supernova bounds on new scalars from
  resonant and soft emission}},
  \href{https://arxiv.org/abs/2410.17347}{{\ttfamily 2410.17347}}.

\bibitem{Giannotti:2015kwo}
M.~Giannotti, I.~Irastorza, J.~Redondo and A.~Ringwald, \emph{{Cool WISPs for
  stellar cooling excesses}},
  \href{https://doi.org/10.1088/1475-7516/2016/05/057}{\emph{JCAP} {\bfseries
  05} (2016) 057} [\href{https://arxiv.org/abs/1512.08108}{{\ttfamily
  1512.08108}}].

\end{thebibliography}\endgroup
\bibliographystyle{jhep}

\end{document}